\documentclass[11pt,usletter]{article}
\usepackage{jheppub}

\title{Effective field theory in time-dependent settings}
\author{Hael Collins, }
\author{R.~Holman, }
\author{Andreas Ross}
\affiliation{Physics Department, Carnegie Mellon University, Pittsburgh PA 15213 USA}
\emailAdd{hcollins@andrew.cmu.edu}
\emailAdd{rh4a@andrew.cmu.edu}
\emailAdd{andreasr@andrew.cmu.edu}

\abstract{
We use the in-in or Schwinger-Keldysh formalism to explore the construction and interpretation of effective field theories for time-dependent systems evolving out of equilibrium.  Starting with a simple model consisting of a heavy and a light scalar field taken to be in their free vacuum states at a finite initial time, we study the effects from the heavy field on the dynamics of the light field by analyzing the equation of motion for the expectation value of the light background field.  New terms appear which cannot arise from a local action of an effective field theory in terms of the light field, though they disappear in the adiabatic limit.  We discuss the origins of these terms as well as their possible implications for time dependent situations such as inflation.}

\keywords{}
\arxivnumber{}

\begin{document}
\maketitle

\setcounter{page}{2}
\section{Introduction}
\label{sec:Intro}

Effective theories provide a powerful method for understanding nature.  Their basic idea is to include only what is relevant for the particular phenomenon that is being described.  For example, to understand how a set of particles scatter and interact, it is not necessary to find a description that applies to all scales, but only one that works for those energies and momenta and precisions that can actually be measured.  All of the further details of nature at higher energies, or shorter distances, can be treated as though they act at a point from the perspective of what a lower energy experiment can resolve.  The symmetries, both of space-time and of the particles' interactions, are an important part of this picture.  In most particle physics applications, the set of symmetries is quite large, and quite constraining, and usually assumes an invariance under the Poincar\'e transformations.  But other settings could have less symmetry.  In the very early universe, the space-time expansion introduces an explicit time-dependence into the action of any quantum fields present.  Nevertheless, the effective theory idea is a universal one which is not restricted to a specific set of symmetries.  As long as a clear separation of the natural dynamical scales exists within a system, and as long as the theory is being applied to measurements at energies well below the higher of the scales, an effective description should exist even when there is an explicit time-dependence.

Before explaining what is new for such settings, it is important to remember how effective field theories are ordinarily applied to scattering experiments \cite{Georgi:1994qn,Rothstein:2003mp,Burgess:2007pt}.  A quantum field is introduced for each particle whose mass is below the maximum energy available to an experiment, and all of the symmetries are identified.  These symmetries typically include the set of Poincar\'e transformations plus whatever other symmetries might be relevant---gauge symmetries, chiral symmetries, permutation symmetries, etc.  The action for the effective theory is then given by the set of all the local operators built from these fields that are invariant under the given symmetries.  This prescription generates an infinite set of operators, though the fact that any given experimental measurement is of finite precision means that in practice only a finite number of them are needed, as long as the energies and momenta of the particles are not too large.

An effective field theory of this sort is never meant to be a fundamental theory, final and absolute, but only one that is appropriate up to a limit.  This limit is usually expressed as a mass scale $M$, and the theory remains predictive for processes whose energies and momenta are below $M$.  But this scale has a deeper meaning too.  Any effective theory always recognizes the possibility that there are further heavier dynamics which cannot be probed directly at low energies.  So above this scale $M$, there should be new fields and interactions, whose masses were too heavy to have had been seen at lower energies.  This higher energy theory is not a final theory either.  Once we have identified all of its fields and symmetries, the game begins anew:  we construct another effective theory which is applicable up to some still higher mass scale where some yet further dynamics remain to be discovered.

If we know the deeper theory or if we have a conjecture for what it might be, we can use it to explain the origin of the low energy effective theory.  Both theories should make the same predictions for the same physical processes.  If we evaluate a process in the higher energy theory, in which external light particles are exchanging virtual heavy particles, the low energy effective theory must yield the same result even though it has no heavy particles to exchange.  This matching of the predictions of the two theories is accomplished by including the appropriate set of interactions in the effective theory.  They have the same external structures in light fields, but unlike the deeper theory they are without any internal structure.  As Feynman graphs, lines representing a heavy field in the higher energy theory are shrunk to a point in the corresponding graphs of the effective theory.

The fact that an action composed of local operators of the relevant light fields can describe any low energy experiment is an important insight underlying effective field theories. It is rooted in the decoupling theorem \cite{Appelquist:1974tg}, which states that any heavy fields that we have left out of our effective theory will influence the dynamics at low energies only indirectly, by inducing or renormalizing the local operators of the light fields.

The processes that are produced by graphs containing internal heavier particles in the high energy theory are reproduced by an infinite set of local operators in the effective theory.  Seen thus, the coefficients of all of the local operators in the effective theory are completely fixed by the parameters of the higher energy theory.  This relation means that even before we have sufficient energies to probe its dynamics directly, we can infer something about the deeper theory from the sizes and relations amongst the coefficients of the effective theory.

This in brief is the picture for how effective theories are used to analyze particle scattering experiments.  There the relevant physical quantity is the amplitude for one set of initial particles to pass into another set of final particles, which is described by the scattering matrix, or $S$-matrix.  The detailed time-evolution is not so important as the asymptotic inputs and results, and the action is invariant under time-translations as a part of a larger Poincar\'e invariance.  While this is an appropriate symmetry for the times and distances over which particle interactions occur in a scattering experiment, in other quantum systems some of the basic space-time symmetries are explicitly broken.

An important example is the theory of inflation.  In inflation, the field responsible for an accelerated expansion of the universe also undergoes quantum fluctuations.  These fluctuations are an essential part of the picture.  They are meant to provide the initial spatial inhomogeneities in the universe that eventually grow into all of the structures that we see.  But unlike particle theories, the action for this quantum field contains an explicit dependence on time inherited from the expanding background.  The physical observables that we are calculating are different too:  we should like to know how the expectation values of products of the fluctuations evolve from some initial state, rather than how initial and final states overlap as in a scattering experiment.

An explicit time-dependence is a generic property of any quantum system that is not in an equilibrium state.  In a condensed matter system, for example, we could raise or lower the temperature, subject it to external fields, and then watch how it responds or relaxes.  Even in particle physics, in the theory of baryogenesis for example, sometimes a stage of out of equilibrium evolution is an essential ingredient.

This article explores how the effective theory idea can be applied to time-dependent systems.  More precisely, we shall try to understand how to treat systems that start in a simple nonequilibrium state and evolve in time.  Since we are investigating time-dependent settings, typically with initial times and finite evolutions, we shall be following the time-evolution of the expectation values of observables rather than calculating scattering processes using the $S$-matrix.  In some settings without Poincar\'e invariance, as in the case of quantum fields in de Sitter space for instance, there is not even a well defined $S$-matrix \cite{Witten:2001kn}, clearly indicating the shortcomings of this observable in time-dependent situations.  The appropriate formalism for following the evolution in a time-dependent setting was first developed by Schwinger and Keldysh \cite{Schwinger:1960qe,Keldysh:1964ud,Bakshi:1962dv,Bakshi:1963bn}.  It allows us to choose arbitrary initial states and it automatically maintains a consistently causal evolution of any observable quantity.  While these methods have been often applied to nonequilibrium quantum systems (see for example \cite{2004cond.mat.12296K}), and have more recently begun to be more widely used for inflationary models, they are perhaps less familiar to many effective field theorists, so in the next section we review the Schwinger-Keldysh formalism.

Most of this article studies a particular example of a time-dependent theory consisting of a  single light field and a single heavy field playing the roles of a heavy and light sector.  These fields interact with each other and, more importantly, their action includes an explicit dependence on the initial time.  An effective theory, even a time-dependent one, must reproduce the same physical predictions at low energies as the higher energy theory.  So as a minimal prescription we demand that at energies well below the scales associated with the heavy field, 
$$
\left\{ { \hbox{expectation values of operators}
\atop\hbox{in the full theory}} \right\}
= 
\left\{ { \hbox{expectation values of operators}
\atop\hbox{in the effective theory}} \right\} , 
$$
where the operators are made up only of the light field.  This matching condition can be taken to define what we mean by an effective theory, given a particular higher energy theory.

One reason for examining a system with a particular time-dependence, rather than considering the most general case from the start, is that once we have broken the time-translation invariance, we could have, in principle, arbitrarily time-dependent structures in the action of the effective theory.  By restricting to a specific case, we can learn whether certain forms of time-dependence translate into generic structures in the low energy theory---or what an effective description even means in this context.  For the simple system that we have investigated, we have found that the effective description contains in part all of the local Poincar\'e invariant operators consistent with a few additional symmetries, and in part some nonlocal transient operators too, related to the initial state that we chose.  At leading nontrivial order in the strength of the coupling between the two fields, these two types of operators are sufficient to describe the effective theory.  Moreover, the transient operators have a universal time-dependence in their leading rate of decay.  This means that with sufficient time, these operators decay to a level below the threshold that would be detectable for a particular experimental precision, and we are left at this order with only the same local action that would have been found in an $S$-matrix derivation of the effective theory.  But at higher orders, we have found that other effects also appear in the expectation values evaluated in the full theory, which are more difficult to express as having arisen from operators in an action for the effective theory. 

In the next section we summarize the tools needed for our analysis:  the Schwinger-Keldysh formalism and the tadpole method used to construct the equation of motion of the expectation value of a field.  We then apply these to our model and use the effective equation of motion for the expectation value of the light field to infer the types of terms needed in the effective action to reproduce this equation.  We conclude with a discussion of the reasons for the appearance of the nonlocal terms and point out further applications of our results.

\section{The Schwinger-Keldysh formalism and the tadpole method}
\label{sec:prelim}

\subsection{The Schwinger-Keldysh formalism}

The observables of interest to us are the time-dependent expectation values of operators ${\cal O}(t)$, 
\begin{equation}
{\rm tr}\bigl( \rho(t){\cal O}(t) \bigr) ,
\label{eq:expect}
\end{equation}
where $\rho(t)$ is the density matrix describing the system.  For most situations we can take $\rho(t)$ to describe a pure state such as the vacuum state of a free field theory, in which case $\rho(t)=|0(t)\rangle\langle 0(t)|$ and where 
\begin{equation}
{\rm tr}\bigl( \rho(t){\cal O}(t) \bigr) = \langle 0(t)|{\cal O}(t)|0(t)\rangle,
\label{eq:expect}
\end{equation}
but we leave open the possibility that the state may actually be mixed, as for a thermal state.  For practical calculations, it is often convenient to evaluate the time-evolution in the interaction picture.  If we separate the Hamiltonian into its free and interacting parts, $H=H_0+H_I$, then in the interaction picture, the time-evolution of the operators is given by the free part, while the evolution of the density matrix is given by the interacting part.  The density matrix satisfies the Liouville equation, 
\begin{equation}
i {\partial\rho(t)\over\partial t} = [H_I(t),\rho(t)],
\label{eq:liouville}
\end{equation}
where $H_I(t)$ is the interaction Hamiltonian in the interaction picture. 

Given an initial state $\rho(t_0)$, the Liouville equation can be solved in terms of the time evolution operator $U_I(t,t_0)$ satisfying the usual Dyson equation,
\begin{equation}
i{\partial U_I(t,t_0)\over\partial t} = H_I U_I(t, t_0) 
\label{eq:liouvillesoln}
\end{equation}
where $U_I(t_0, t_0)=\mathbb{I}$ and 
\begin{equation}
\rho(t) = U_I(t,t_0) \rho(t_0) U_I^\dagger(t,t_0) . 
\label{eq:liouvillesoln}
\end{equation}
The solution that satisfies this initial condition is 
\begin{equation}
U_I(t,t_0) = T e^{-i\int_{t_0}^t dt'\, H_I(t')} .
\end{equation}
In the interaction picture then, the expectation value of an operator is given by 
\begin{eqnarray}
{\rm tr}\bigl( \rho(t){\cal O}(t) \bigr) 
&=& {\rm tr}\bigl[ \rho(t_0) U_I^\dagger(t,t_0) {\cal O}(t) U_I(t,t_0) \bigr] 
\nonumber \\
&=& {\rm tr}\Bigl[ \rho(t_0)\Bigl( T e^{-i\int_{t_0}^t dt'\, H_I(t')}\Bigr)^\dagger 
{\cal O}(t) \Bigl( T e^{-i\int_{t_0}^t dt'\, H_I(t')}\Bigr) \Bigr] .
\label{eq:ExpValint}
\end{eqnarray}
Reading the terms inside the trace from right to left, the first operator evolves from $t_0$ to $t$, where the operator ${\cal O}(t)$ is inserted, and then we turn around and evolve back to $t_0$ again.  Because of the cyclicity of the trace, we are starting and ending in the same state, which is defined by $\rho(t_0)$.

We can use this way of viewing the time evolution to define a contour time-ordered product.  Let us introduce $\pm$ labels for the fields, where a $+$ field occurs on the on the forward part of this contour, running from $t_0$ to $t$, and a $-$ field occurs on the backward part that runs from $t$ back to $t_0$ again.  The $+$ fields are then associated with the operator $U_I(t,t_0)$ while the $-$ fields are associated with the $U_I^\dagger(t,t_0)$ operator.  We can extend the time integral into the infinite future as well by judiciously inserting a factor of the identity, ${\mathbb I} = U_I^\dagger(\infty,t)U_I(\infty,t)$ between the $U^\dagger_I(t,t_0)$ and ${\cal O}(t)$ in the previous equation, writing\footnote{We typically assume that the operator being evaluated is placed on the $+$ part of the contour.  However, evaluating it on the $-$ contour would give exactly the same result.}
\begin{equation}
{\rm tr}\bigl( \rho(t){\cal O}(t) \bigr) 
= {\rm tr} \Bigl[ \rho(t_0) T_c \Bigl( {\cal O}^+(t) 
e^{-i\int_{t_0}^\infty dt'\, [H_I^+(t')-H_I^-(t')] }\Bigr)  \Bigr]. 
\end{equation}
The $H^\pm_I$ are of exactly the same form as the original interacting part of the Hamiltonian, except that in $H^+_I$ we label all the fields within it as $+$ fields (e.g.~$H_I^+(t)=H_I[\Phi^+(t,\vec x)]$) and in $H_I^-$ they are all labeled as $-$ fields.  The subscript on the time-ordering symbol $T_c$ indicates that the time-ordering is that inherited from the closed time contour:   times that occur in a $-$ field always occur {\it after\/} and {\it in the opposite order\/} of those in a $+$ field.  This follows from the original Hermitian conjugation $U_I^\dagger(t,t_0)$ and the fact that it occurs furthest to the left---the `latest' in the contour sense---in the matrix element.

When $H_I$ is small in comparison with the free Hamiltonian, we can evaluate the expectation value perturbatively, expanding in terms of propagators of the free theory.  In the interaction picture, the state does not evolve for a free theory, so the propagator is evaluated using the initial state, $\rho(t_0)$.  There are four possible ways of contracting the field depending on whether each of the two fields is on the forward ($+$) or backward ($-$) part of the contour.  This means that in any process, we have four possible time-ordered Green's functions, 
\begin{equation}
G^{\pm\pm}(t,\vec x;t',\vec y) 
= {\rm tr}\left(\rho(t_0) T_c\bigl( \Phi^\pm(t,\vec x)\Phi^\pm(t',\vec y) \bigr)\right). 
\end{equation}
Using the definition of the time-ordering along the contour, these Green's functions are 
\begin{eqnarray}
G^{++}(t,\vec x; t',\vec y) &=& 
\Theta(t-t')\, G^>(t,\vec x; t',\vec y) 
+ \Theta(t'-t)\, G^<(t,\vec x; t',\vec y) 
\nonumber \\
G^{+-}(t,\vec x; t',\vec y) &=& G^<(t,\vec x; t',\vec y)
\nonumber\\
G^{-+}(t,\vec x; t',\vec y) &=& G^>(t,\vec x; t',\vec y)  
\nonumber \\
G^{--}(t,\vec x; t',\vec y) &=&
\Theta(t'-t)\, G^>(t,\vec x; t',\vec y) 
+ \Theta(t-t')\, G^<(t,\vec x; t',\vec y)
\label{eq:greenshom}
\end{eqnarray}
where the two Wightman functions are given by the two possible orderings of the fields depending on whether $t>t'$ or $t<t'$, 
$$
G^>(t,\vec x; t',\vec y) =
{\rm tr}\left(\rho(t_0) \Phi(t,\vec x)\Phi(t',\vec y)\right)
\qquad\hbox{and}\qquad 
G^<(t,\vec x; t',\vec y) = 
{\rm tr}\left(\rho(t_0)\Phi(t',\vec y)\Phi(t,\vec x)\right).
$$

We shall mostly be examining the case where $\rho(t_0)$ is the free field vacuum state, which we shall write as $|0(t_0)\rangle\equiv|0\rangle$.  Since this state is invariant under spatial translations, we can conveniently express these Wightman functions in their Fourier transformed form,
\begin{eqnarray}
G^>_k(t,t') &=& {1\over 2\omega_k} e^{-i\omega_k(t-t')}
\nonumber\\
G^<_k(t,t') &=& {1\over 2\omega_k} e^{i\omega_k(t-t')},
\label{eq:freeprops}
\end{eqnarray}
where $k \equiv |\!|\vec k|\!|$ is the magnitude of the spatial momentum and $\omega_k \equiv \sqrt{k^2+m^2}$. 

For many purposes it is useful to represent the expectation value as a path integral,
\begin{eqnarray}
{\rm tr}\bigl( \rho(t){\cal O}(t) \bigr)
&=&
\int d\phi\, \langle\phi| \rho(t_0) U_I^\dagger(t,t_0){\cal O}(t) U_I(t,t_0) 
|\phi\rangle 
\label{eq:ctpexpect} \\
&=& 
\int d\phi^+d\phi^-\,  \rho(\phi^+,\phi^-;t_0)  \int_{\rm bc} {\cal D}\Phi^+{\cal D}\Phi^-\, e^{i(S[\Phi^+]-S[\Phi^-])} {\cal O}(t) ,
\nonumber
\end{eqnarray}
where we have first used cyclicity of the trace and then evaluated it in the field basis $\left\{|\phi\rangle\right\}$.  Just as before, we can interpret this trace as follows:  start at $t_0$ with the initial state, evolve to $t$ where the operator ${\cal O}(t)$ is inserted and then evolve back to $t_0$.  The boundary conditions on the path integrals over $\Phi^\pm$ are $\Phi^+(t_0,\vec x) = \phi^+(\vec x)$, $\Phi^-(t_0,\vec x) = \phi^-(\vec x)$, and $\Phi^+(t,\vec x) = \Phi^-(t,\vec x)$.  The last condition can be interpreted as a continuity condition for a field $\Phi_c$ defined along a time contour starting at $t_0$, going to $t$, turning around and then returning to $t_0$.  This is the origin of the closed time contour and the difference in actions in the exponential is just the integral of the Lagrangian density along this closed time contour, taking the orientation of the contour into account. 

We can define a generating functional from this path integral by introducing sources $J^\pm$ which couple linearly to the $\Phi^\pm(t,\vec x)$ fields,
\begin{equation}
Z[J^+, J^-] = \int d\phi^+d\phi^-\,  
\rho(\phi^+,\phi^-;t_0)  
\int_{\rm bc} {\cal D}\Phi^+{\cal D}\Phi^-\, 
e^{i(S[\Phi^+\!\!,\, J^+] - S[\Phi^-\!\!,\, J^-])} .
\label{eq:genfunct}
\end{equation}
As before, we can always extend the time integrals, which are implicit in the actions, 
\begin{equation}
S[\Phi^\pm\!\!,\, J^\pm] = \int_{t_0}^t dt' \int d^3\vec x\, \Bigl\{ 
{\cal L}[\Phi^\pm(t',\vec x)] + J^\pm(t',\vec x)\Phi^\pm(t',\vec x) \Bigr\} , 
\end{equation}
into the infinite future since the forward and the backward parts of the contour beyond $t$ (the time at which we are evaluating a particular operator) exactly cancel---as they must if the theory is to remain causal.  

Notice that there are two marked differences between this closed time path formalism and the more commonly used $S$-matrix treatment.  In a scattering process, we start in the infinite past with an eigenstate of the free theory, evolve it into the infinite future, and find its overlap with the eigenstates of the free theory again.  In that case, there is only one occurrence of the time-evolution operator, $U_I(\infty,-\infty)$.  In the Schwinger-Keldysh approach, in contrast, we are evolving the entire matrix element forward, evolving both states, and accordingly both $U_I^\dagger$ and $U_I$ appear.  

The second difference is that we are evolving over a finite interval and from a general initial state defined at an arbitrary initial time.  We are free to consider a situation where $t_0\to -\infty$ or to choose the system to be the vacuum state of the free theory, but whatever state we have chosen, once we have specified its value at $t_0$, the preceding history is irrelevant for its subsequent evolution.  The lower limit of the integrals in the exponents of the time-evolution operators is always $t_0$, so any intermediate vertices or any loop corrections are never evaluated prior to the initial time.

In some instances, it can be useful to extend the integral formally into the infinite past.  Whether this should be done depends on the detailed physical situation and initial state that we are studying.  For example, if we arrange the interactions prior to $t_0$ so that our system will be in precisely the desired state by that time, then we can extend the time integral in these actions to cover the entire real line.  To arrange the state thus will almost always require an explicit time-dependence in the action---in the couplings for instance---if we do not happen to start with an eigenstate of the interacting theory.

For most of this article we shall be examining a system that begins in its free vacuum state at $t_0$ and then evolves under the influence of some interactions.  This could be arranged by multiplying the interactions in $H_I$ by a step-function, $\Theta(t-t_0)$, so that if the theory has been evolving freely before $t_0$ it will have the behavior that we wish afterwards.  This perspective is heuristically useful since it shows that by starting in the vacuum state of the free theory rather than the interacting one we can expect some interesting evolution, since the $\Theta$-function represents a very sudden change to the system.  But ultimately the details of how the system arrived in a particular state at $t_0$ are irrelevant for its subsequent evolution.

\subsection{The tadpole method}
\label{subsec:tadpole} 

One practical difference between calculating a matrix element in the Schwinger-Keldysh and the $S$-matrix frameworks is that it is usually not possible to amputate the external legs of a process in the former approach.  While this fact is not in itself anything profound, it does mean that the calculations of processes in the Schwinger-Keldysh picture will typically be lengthier than those for the analogous graphs in a scattering process.  It is therefore helpful to have tools or tricks that can simplify the analysis whenever possible.  One such tool is the tadpole method \cite{Weinberg:1973ua,Boyanovsky:1994me}, which uses the equation of motion for a classical background field to follow the quantum loop corrections of a theory.

Consider a quantum field $\Phi(t,\vec x)$ that has a time-dependent expectation value when it is in a state specified by the density matrix $\rho(t)$, 
\begin{equation}
\phi(t) = {\rm tr}\bigl( \rho(t)\Phi(t,\vec x)\bigr) .
\end{equation}
In most models of inflation, it is just such a classical expectation value of a field which drives the accelerated cosmological expansion, while the quantum fluctuations about this background provide a source for all of the subsequent inhomogeneity of the universe.  Condensed matter systems that undergo a phase transition can similarly have order parameters that evolve in time.  

Setting a constant one-point function, or tadpole diagram, to zero in a field theory ensures that the quantum field is an excitation above its correct vacuum state; this condition is equivalent to calculating the amount by which we must shift the field such that we expand about the correct vacuum expectation value.  The analogue of this statement is also true for non-constant backgrounds, where setting the one-point function to zero ensures that the background satisfies the correct equation of motion.

When we evaluate an expectation value in the Schwinger-Keldysh picture, we expand about the same classical background $\phi(t)$ on both branches of the time-contour, 
$$
\Phi^\pm(t,\vec x) = \phi(t) + \varphi^\pm(t,\vec x) , 
$$
where $\varphi^\pm(t,\vec x)$ is the part of the quantum field representing the fluctuations about the classical part.  By doing so, we arrive at the same equation of motion for $\phi(t)$ regardless of whether we place $\varphi(t,\vec x)$ on either part of the contour when evaluating its one-point function, 
$$
{\rm tr}\bigl( \rho(t)\varphi^\pm(t,\vec x)\bigr) = 0 . 
$$

Notice that although the one-point function for $\varphi^\pm(t,\vec x)$ has only one external leg for the fluctuations, it can have arbitrarily many external (nonpropagating) legs for the classical field $\phi(t)$.\footnote{We shall refer to the appearances of the classical field $\phi(t)$ as legs of a diagram---and we shall draw them as external dotted lines in graphs---though of course $\phi(t)$ is not a propagating field.  We adopt this language and use this notation nonetheless since through it the graphs more closely resemble the corresponding $N$-point processes for which they are the analogues.}
\begin{figure}[htbp]
\begin{center}
\includegraphics{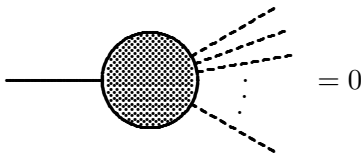}
\caption{{The diagrammatic form of the tadpole condition.  The left side of the equation consists of all diagrams with one external leg for the quantum field $\varphi(t,\vec x)$, which is represented here by the single solid line.  The diagram can have an arbitrary number of external legs corresponding to the background field $\phi(t)$, which are shown here as dashed lines.  At any finite order in a perturbative expansion in the coupling, only a finite number of diagrams contribute.}}
\label{generaltadpole}
\end{center}
\end{figure}
The requirement that the sum of all the tadpole graphs must vanish produces an equation for the background field $\phi(t)$.  This equation contains the loop corrections from the quantum part of the field so it reproduces most of the familiar renormalization properties of the theory.  It therefore provides a simpler route for analyzing the structures of the effective theory than is possible by evaluating the higher-order $N$-point functions of $\Phi^\pm(t,\vec x)$ directly.

At lowest order, the tadpole condition is just the statement that the field $\phi(t)$ satisfies its classical equation of motion.  To show this, write the action for the field on each of the two parts of the contour as $S^\pm[\Phi^\pm]$ and expand the field about its background value, $\Phi^\pm(t',\vec y) = \phi(t')$. Keeping only the linear terms, we find
\begin{equation}
S^\pm[\Phi^\pm] = S^\pm[\phi] + \int_{t_0}^\infty dt'\int d^3\vec y\, \biggl\{ 
{\delta S^\pm(\Phi^\pm)\over\delta\Phi^\pm} 
\biggr|_{\Phi^\pm(t',\vec y)=\phi(t')} \varphi^\pm(t',\vec y) \biggr\} 
+ {\cal O}({\varphi^\pm}^2) . 
\end{equation}
Since $S^+[\phi]=S^-[\phi]$, the zeroth order terms cancel in the difference that appears in the generating functional.  At linear order in both the expansion of the action and the exponentials, the one-point function of $\varphi(t,\vec x)$ is then 
\begin{eqnarray}
&&\!\!\!\!\!\!\!\!\!
{\rm tr}\bigl( \rho(t)\varphi^+(t,\vec x) \bigr) 
\label{eq:treetad} \\
&=& \int {\cal D}\varphi^+{\cal D}\varphi^-\, \biggl\{ \varphi^+(t,\vec x) 
e^{i\int_{t_0}^\infty \int dt'd^3\vec y\, \left[ 
{\delta S^+(\Phi^+)\over\delta \Phi^+(t',\vec y)} \bigr|_{\phi(t')} \varphi^+(t',\vec y) 
- {\delta S^-(\Phi^-)\over\delta \Phi^-(t',\vec y)} \bigr|_{\phi(t')} \varphi^-(t',\vec y) \right]
\ +\ {\cal O}(\varphi^{\pm 2}) }
\biggr\}
\nonumber \\
&=& 
i\int_{t_0}^\infty dt' \int d^3\vec y\, \biggl\{ 
{\delta S^+\over\delta\Phi^+(t', \vec y)}\biggr|_{\phi(t')} 
G^{++}(t,\vec x;t',\vec y) 
- {\delta S^-\over\delta\Phi^-(t',\vec y)}\biggr|_{\phi(t')} 
G^{+-}(t,\vec x;t',\vec y) 
\biggr\} + \cdots .
\nonumber 
\end{eqnarray}
The Green's functions for the fluctuations that appear in these two terms are independent so that for the one-point function to vanish, this integrand must  also vanish, which in turn requires that 
\begin{equation}
{\delta S^\pm\over\delta\Phi^\pm}\biggr|_{\Phi^\pm=\phi(t)} = 0 .
\end{equation}
This condition is nothing more than the statement that the action is stationary at $\phi(t)$ to leading order---precisely the condition that defines the classical equation of motion for $\phi(t)$.

What makes the tadpole condition such a useful tool is that it also incorporates all of the higher order operators of $\varphi^\pm(t,\vec x)$ once we have proceeded beyond linear order.  Any product of operators with an odd number of $\varphi^\pm$ fields, when contracted with the external $\varphi^+(t,\vec x)$ being evaluated, will produce loop corrections to the equation of motion for the background field $\phi(t)$.  This procedure can be best illustrated through a familiar example.

Consider a scalar field with a quartic self-interaction, 
\begin{equation}
S_\Phi = \int dtd^3\vec x\, 
\Bigl\{ {\textstyle{1\over 2}} \partial_\mu\Phi\partial^\mu\Phi 
- {\textstyle{1\over 2}} m^2 \Phi^2 
- {\textstyle{1\over 24}} \lambda \Phi^4 \Bigr\}.
\end{equation}
If we expand the field about its expectation value, $\Phi(t,\vec x) = \phi(t) + \varphi^\pm(t,\vec x)$, and gather the operators according to their powers of the quantum fluctuations, we find 
\begin{eqnarray}
S^+_\Phi - S^-_\Phi &=& 
\int dtd^3\vec x\, \Bigl\{ 
{\textstyle{1\over 2}} \partial_\mu\varphi^+\partial^\mu\varphi^+
- {\textstyle{1\over 2}} m^2\varphi^{+2} 
- \varphi^+ \Bigl[ \ddot\phi + m^2\phi 
+ {\textstyle{1\over 6}} \lambda \phi^3 \Bigr]
\nonumber \\
&&\qquad\quad\ \,
-\, {\textstyle{1\over 4}} \lambda\, \varphi^{+2} \phi^2
- {\textstyle{1\over 6}} \lambda\, \varphi^{+3} \phi 
- {\textstyle{1\over 24}} \lambda\, \varphi^{+4} 
- (\varphi^+\leftrightarrow \varphi^-) \Bigr\}.
\end{eqnarray}
The first two terms correspond to the free Lagrangian for the field $\varphi^+(t,\vec x)$.  From the perspective of these fluctuations, the rest of the terms represent a set of interactions with the background $\phi(t)$. Notice that we have chosen to treat the ${1\over 2}\lambda\phi^2\varphi^{+2}$ as an interaction, even though it is quadratic in $\varphi^+(t,\vec x)$.  As long as ${1\over 2}\lambda\phi^2\ll m^2$, we can do so consistently; when this condition is not satisfied, we must instead define an effective, time dependent mass, $m^2_{\rm eff}(t) = m^2+{1\over 2}\lambda\phi^2(t)$.  The behavior of the field in this regime can then be found numerically \cite{Boyanovsky:1994me}.

\begin{figure}[htbp]
\begin{center}
\includegraphics{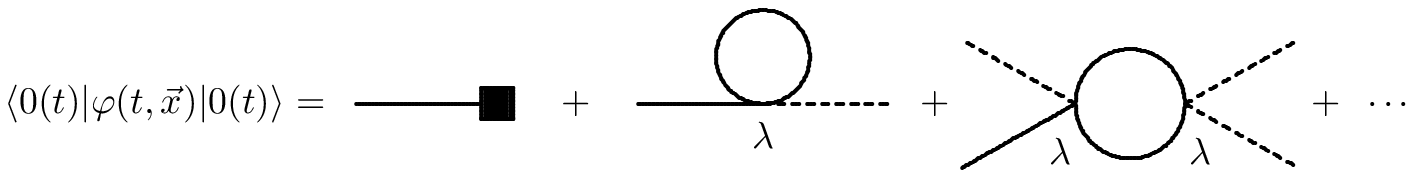}
\caption{{The graphs contributing to the zero mode equation of motion.  A solid line again corresponds to the quantum fluctuation, $\varphi(t,\vec x)$, while the dashed lines indicate the classical background $\phi(t)$.  In the first graph, the shaded square corresponds to the tree-level equation of motion operator $\left(\square^2+m^2\right)\phi + {1\over 6} \lambda\phi^3$, though we have not shown the $\phi(t)$ lines attached to it.}}
\label{fig:phi4eom}
\end{center}
\end{figure}

For simplicity, let us choose the field to be in its free vacuum state initially, at $t=t_0$, $\rho(t_0) = |0(t_0)\rangle\langle 0(t_0)|$, with its subsequent evolution given by the interaction terms in the action, $|0(t)\rangle = U_I(t,t_0)|0(t_0)\rangle$.  If we evaluate the tadpole condition for this state to second order in $\lambda$, then we generate the set of graphs shown in figure \ref{fig:phi4eom}.
The structure of the one-point function to this order is
\begin{equation}
\langle 0(t_1)|\varphi(t_1,\vec x)|0(t_1)\rangle 
= - \int_{t_0}^{t_1} dt\, {\sin[m(t_1-t)]\over m}\, 
\Bigl\{ \cdots \Bigr\} 
= 0 ;
\end{equation}
the first factor in the integrand corresponds to the $\varphi$-propagator, with no spatial momentum flowing through it, and the ellipses corresponds to an equation of motion for the background field $\phi(t)$, 
\begin{eqnarray}
&& 
\ddot\phi(t) + m^2\phi(t) + {\lambda\over 6} \phi^3(t)
+ {\lambda\over 2} \phi(t) \int{d^3\vec k\over (2\pi)^3}\, {1\over 2\omega_k} 
\nonumber \\
&&\qquad 
- {\lambda^2\over 4} \phi(t) \int^t_{t_0}dt'\, \phi^2(t')
\int {d^3\vec k\over (2\pi)^3}\, 
{\sin\left[2\omega_k(t-t^{\prime})\right]\over 2\omega^2_k} + {\cal O}(\lambda^3) = 0 . \qquad
\label{eq:phi4eom}
\end{eqnarray}
The first three terms are just an instance of the general result that we found before, that the tree-level terms correspond to the classical equation of motion for $\phi(t)$.  The first of the integral terms---and the second of the graphs in the figure---is a quadratically divergent loop integral.  It is linear in $\phi(t)$, and it can be recognized as the familiar divergent correction to the mass, which can be absorbed by defining a renormalized mass for the field.

The last of the terms in the equation of motion, represented by the last of the graphs, is cubic in the background field, although two of the fields are evaluated at the intermediate time $t'$ over which we are integrating.  The integral over the spatial loop momentum is divergent, though the sine function obscures the precise form of this divergence somewhat since it vanishes at $t=t'$.  To isolate the divergent part of this graph, let us define the kernel, 
\begin{equation}
K(t-t') = \int {d^3\vec k\over (2\pi)^3}\, {\cos[2\omega_k(t-t')]\over\omega_k^3} .
\label{eq:kernel}
\end{equation}
This function clearly has a logarithmic divergence for large values of $|\!|\vec k|\!|$ when $t=t'$.  Its $t'$ derivative is essentially the loop integral that appears in the cubic correction to the equation of motion for the background field $\phi(t)$.  We use this property to integrate it by parts to produce three terms
\begin{eqnarray}
\int_{t_0}^t dt'\, \phi^2(t') \int {d^3\vec k\over (2\pi)^3}\, 
{\sin[2\omega_k(t-t')]\over 2\omega^2_k} 
&=& {1\over 4} \int_{t_0}^t dt'\, \phi^2(t') {d\over dt'} K(t-t') 
\nonumber \\
&=& 
{1\over 4} \phi^2(t) K(0) - {1\over 4} \phi^2(t_0)K(t-t_0) 
\nonumber \\
&&
- {1\over 2} \int_{t_0}^t dt'\, \phi(t') \dot\phi(t') K(t-t') . 
\label{eq:quarticrenorm}
\end{eqnarray}
Multiplying by the ${1\over 4}\lambda^2\phi(t)$ factor that appeared in the equation of motion, the first of these terms becomes ${1\over 8}\lambda^2 K(0)\, \phi^3(t)$.  It is the standard logarithmically divergent one-loop correction to the quartic coupling, which can be absorbed by a renormalization of $\lambda$.  The final term in the equation includes dissipative effects, as is shown by the $\dot\phi(t')$ within the integral.  It occurs because the nonlinear coupling allows the background to transfer its energy into excitations of the quantum fluctuations.

The second term, once we have included the necessary prefactor, has the form
\begin{equation}
\textstyle
- {1\over 16}\lambda^2\phi^2(t_0)K(t-t_0)\, \phi(t) . 
\end{equation}
It is proportional to $\phi(t)$, which might seem to suggest that it is a mass term, albeit one with a time-dependent mass since the kernel depends on time.  However, the field itself is also evaluated on the boundary, as shown by the factor $\phi^2(t_0)$, which indicates that this term is actually associated with a quartic operator at the level of the full action for the original quantum field $\Phi(t,\vec x)$.  As we shall show later, the amplitude of the kernel decays as its argument grows large, $K(t-t_0)\to 0$ as $t-t_0\to\infty$.  

If we had prepared the system in the infinite past, $t_0\to -\infty$, then this term would vanish.  Alternatively, if we had started with a free theory in the infinite past and turned on the interactions adiabatically, then this term would also never have appeared.  The reason that the time-dependent terms have appeared is that we began evolving our system at $t_0$ from an initial state, the free vacuum, that is not an eigenstate of the interacting theory.  This can alternatively be viewed as evolving the free theory starting in the infinite past with the free vacuum as the initial state, and then turning on the interactions at $t_0$.  From either perspective, this term cannot be neglected, at least during a transient stage after $t_0$.  We are interested here in studying theories where there is an explicit time-dependence in a setting where there also is a hierarchy of scales, so we shall next investigate how to construct effective field theories in such a setting.

\section{The one-point function and the low energy effective theory}
\label{sec:intheavy}

Effective field theories are usually derived for settings that are invariant under the full set of Poincar\'e transformations.  A theory that has an explicit time-dependence has less symmetry, and the operators that are allowed in the low energy effective theory are accordingly less constrained.  Here we shall be considering systems where the time-translation invariance is broken even at the level of the action of the full theory.  While allowing a more dramatic time dependence lets us treat a wider range of physical systems, it also means that not all of the structures that appear in the effective theory are of a form that would have been anticipated from the more familiar $S$-matrix approach, where the symmetry is unbroken.  In the limit that the time-translation symmetry is restored, the effective theory should assume its standard $S$-matrix form---a requirement that provides a weak condition on the structures that are allowed for the new operators in the effective theory.

Our general philosophy for defining the effective theory in a time-dependent setting is that the expectation values evaluated in the effective theory should match, at low energies, with those of the full theory from which it was to have been derived.  For example, in terms of a very general state described by a density matrix $\rho(t)$ in the higher energy theory, this matching condition becomes
\begin{equation}
{\rm tr}\bigl( \rho_{\rm eff}(t){\cal O}(t) \bigr) 
= {\rm tr}\bigl( \rho(t){\cal O}(t) \bigr) , 
\end{equation}
where ${\cal O}(t)$ is an operator composed only of light degrees of freedom.  The trace on the right side is over all the degrees of freedom, while on the left it is only over the light degrees of freedom.

The form of the effective theory is thus contained in the effective density matrix $\rho_{\rm eff}(t)$.  This condition can therefore be used as a prescription how to construct the effective theory when we already know he higher energy theory.  We proceed in this way since we should like to develop some simple examples of high energy theories and their corresponding effective theories.  Such examples will ultimately help to guide us in constructing time-dependent effective theories {\it ab initio\/}; but to do this, we must acquire some intuition for the sorts of structures that could typically appear.

To have a definite example at hand, let us consider a theory composed of a heavy field $\chi(t,\vec x)$ and a light field $\Phi(t,\vec x)$ and described by an interaction Hamiltonian $H_I[\Phi,\chi]$.  We shall place both fields in their vacuum state initially, $\rho(t_0)=|0\rangle\langle 0|$, where $|0\rangle = |0(t_0)\rangle$.  For this choice, the effective density matrix will initially be in the vacuum state for the light field.  Evaluating the one-point function of the light field, ${\cal O}(t)=\varphi(t,\vec x)$, the matching condition becomes
\begin{equation}
{\rm tr}\bigl( \rho_{\rm eff}(t) \varphi(t,\vec x) \bigr) 
= \Bigl\langle 0\Big| T_c \Bigl( \varphi^+(t,\vec x) 
e^{-i\int_{t_0}^\infty dt'\, 
\{ H_I^+[\Phi^+,\chi^+]-H_I^-[\Phi^-,\chi^-]\} }\Bigr) \Big|0\Bigr\rangle , 
\end{equation}
where $\rho_{\rm eff}(t_0)=|0\rangle\langle 0|$.  We shall see how far we are able to derive $\rho_{\rm eff}(t)$---that is the effective theory---from this condition.

If we had a relatively tame time-dependence, where the only changes are happening on scales that are natural from the perspective of the light degrees of freedom, and if none of the heavy degrees of freedom are excited, then we should expect that the effective theory would have a standard unitary evolution, 
\begin{equation}
\rho_{\rm eff}(t) = U_{\rm eff}(t,t_0)\rho_{\rm eff}(t_0) U_{\rm eff}^\dagger(t,t_0) .
\end{equation}
Here the time-evolution operator for the effective theory could expressed in terms of an effective interaction Hamiltonian, $H_{I,{\rm eff}}[\Phi]$,
\begin{equation}
U_{\rm eff}(t,t_0) = T e^{-i\int_{t_0}^t dt'\, H_{I,{\rm eff}}[\Phi] } .
\end{equation}
In this limit, the matching would be very straightforward, 
\begin{eqnarray}
&&\!\!\!\!\!\!\!\!\!\!\!\!\!\!\!\!\!\!\!\!\!\!\!\!\!\!\!\!
\Bigl\langle 0\Big| T_c \Bigl( \varphi^+(t,\vec x) e^{-i\int_{t_0}^\infty dt'\, 
\{ H_{I,{\rm eff}}^+[\Phi^+]-H_{I,{\rm eff}}^-[\Phi^-]\} }\Bigr) 
\Big|0\Bigr\rangle 
\nonumber \\
&=&
\Bigl\langle 0\Big| T_c \Bigl( \varphi^+(t,\vec x) e^{-i\int_{t_0}^\infty dt'\, 
\{ H_I^+[\Phi^+,\chi^+]-H_I^-[\Phi^-,\chi^-]\} }\Bigr) \Big|0\Bigr\rangle . 
\end{eqnarray}

When changes occur over sufficiently short intervals---at or shorter than those comparable to the natural scales of the heavy field---or when somehow heavier modes can be excited, then there is no longer any reason to expect that such a simple formulation of the effective theory is always possible.  In these cases the system formulated only in terms of the light field remains an open one, and it is less clear just how far an effective description even remains possible.

\subsection{Calculating the one-point function to leading order for a $\chi^2\Phi^2$ coupling}

Let us now examine a concrete example.  Consider a theory where the action for the light $\Phi(t,\vec x)$ and heavy $\chi(t,\vec x)$ fields is given by 
\begin{equation}
S = \int dtd^3\vec x\, \Bigl\{
{\textstyle{1\over 2}} \partial_\mu\Phi\partial^\mu\Phi 
- {\textstyle{1\over 2}} m^2\Phi^2 
+ {\textstyle{1\over 2}} \partial_\mu\chi\partial^\mu\chi
- {\textstyle{1\over 2}} M^2\chi^2 
- {\textstyle{1\over 24}} \lambda\Phi^4 
- {\textstyle{1\over 24}}\tilde\lambda\chi^4 
- {\textstyle{1\over 2}}g\chi^2\Phi^2 \Bigr\} .
\end{equation}
and where there is a clear hierarchy of scales, $M\gg m$.  Aside from their individual actions, the two fields interact through a term which is quadratic in both of the fields, 
\begin{equation}
H_I(t) = \int d^3\vec x\, \Bigl\{ {\textstyle{1\over 2}} g\Phi^2\chi^2 \Bigr\} .
\end{equation}

The time-dependence in this system is fairly subtle.  As before we shall let the light field have a classical expectation value, 
$$
\langle 0(t)|\Phi(t,\vec x)|0(t)\rangle = \phi(t), 
$$
while letting that of the heavy field vanish, $\langle 0(t)|\chi(t,\vec x)|0(t)\rangle = 0$.  The background field $\phi(t)$ is a solution to its own equation of motion and while it has an explicit time-dependence, it does not itself break the symmetry of the original action.  The true origin of the breaking of the time-translation symmetry resides in the fact that the evolution starts at a finite initial time, $t_0$, and in our choice of the states.  Both fields are chosen initially to be in their {\it free\/} vacuum states, which we write as $|0(t_0)\rangle = |0\rangle$, but the subsequent evolution is determined by the full action, which includes the interaction term.  Therefore we do not start the system in the vacuum state of the interacting theory nor in any energy eigenstate of the interacting theory.  We choose the free vacuum as a simple example of a nonequilibrium initial state and expect it to entail some interesting time evolution.  One issue here is that it possibly contains excitations of heavy degrees of freedom or of light degrees of freedom at high momenta.  Their presence would invalidate the usual description as a local effective field theory in terms of the light field and should cause some relaxation phenomenon as the heavy or high momentum degrees of freedom diminish with time.

We could make this time-dependence, which is present in the action through the limits of the integration, into one that appears explicitly in the Lagrangian by writing the interaction term as ${\textstyle{1\over 2}}\Theta(t-t_0)g\chi^2\Phi^2$.  We could then study the equivalent problem where the system is placed in the free vacua for both fields at $t\to -\infty$ and then allowed to evolve freely until $t=t_0$ when they begin to interact.  From this perspective it is clear that the time-dependence is a sudden one, although it may not have initially appeared so.  It should not be surprising therefore when later we find that the effective theory contains operators beyond the usual Poincar\'e invariant set.

To explore the structure of the effective theory, we evaluate the one-point function for the fluctuations of the light field, $\varphi(t,\vec x)$, in the presence of a classical background, $\phi(t)$.  Expanding the interaction term about this background, $\Phi^\pm(t,\vec x) = \phi(t) + \varphi^\pm(t,\vec x)$, we generate the following contributions to the interaction Hamiltonian,\footnote{We are again working in the limit where we can treat the ${1\over 2} g\phi^2\chi^2$ term as a part of the interactions rather than as a part of the mass term of the free part of the theory.  This treatment should be consistent as long as $g\phi^2\ll M^2$, which is a natural requirement since $\phi$ is associated with the light and $M$ is associated with the heavy part of the theory.}
\begin{equation}
H_I^\pm(t) = \int d^3\vec x\, \Bigl\{ 
{\textstyle{1\over 2}} g\phi^2{\chi^\pm}^2 + g \phi{\varphi^\pm}{\chi^\pm}^2 
+ {\textstyle{1\over 2}} g {\varphi^\pm}^2{\chi^\pm}^2 \Bigr\} .
\end{equation}
This interaction produces a series of tadpole graphs which can be arranged according to the number of loops and external $\phi$ legs.  At any given order in the coupling $g$ these graphs lead to a finite number of corrections to the $\phi(t)$ equation of motion.  We evaluate the leading nontrivial one-loop correction completely, showing how it generates a series of terms with time-derivative of $\phi(t)$ which are suppressed by an appropriate power of $1/M$. 

\begin{figure}[htbp]
\begin{center}
\includegraphics[scale=1]{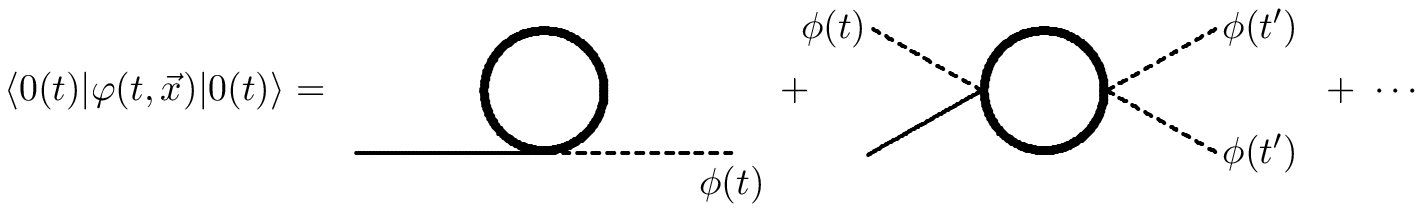}
\caption{Two of the lowest order (in a small coupling expansion) one-loop graphs that contribute to the $\varphi^+$ tadpole condition.  The thicker line corresponds to a $\chi$ field while the thin and dashed lines once again represent the $\varphi$ and $\phi$ fields.}
\label{fig:havyloop}
\end{center}
\end{figure}

The truly leading one-loop contribution to the one-point function of the fluctuation is given by the contraction of the external $\varphi^+(t,\vec x)$ field with the operator $g\phi\varphi^\pm(\chi^\pm)^2$ in the interaction.  This leads to the first graph shown in figure \ref{fig:havyloop}.  It is another quadratically divergent correction to the mass of the light field, which can be removed by defining an appropriate physical mass for it.  We assume that we have done so and consider the order $g^2$, one-loop correction to the equation of motion for the background.  This graph has three external $\phi$-legs as shown in the figure.
Assuming that we have appropriately renormalized the mass of the light field, the equation of motion for $\phi(t)$ assumes the following form when we include the contribution from the second diagram in the figure, 
\begin{equation}
\ddot\phi(t) + m^2\phi(t) + {1\over 6}\lambda \phi^3(t)
- {1\over 2} g^2 \phi(t) \int_{t_0}^t dt'\, \phi^2(t') 
\int {d^3\vec k\over (2\pi)^3}\, 
{\sin\bigl[ 2\omega_k(t-t')\bigr]\over \omega_k^2} + \cdots = 0 .
\end{equation}
Here $\omega_k$ is the once again the energy of the field running around the loop,
$$
\omega_k = \sqrt{k^2+M^2} , 
$$
though this time it is the heavy $\chi$ field that is doing so.

We recognize the structure of the last term from our earlier analysis of the $\lambda\Phi^4$ self-coupling.  A crucial change from before is that now the integral over the loop momentum depends on the mass of the heavy field rather than the light one.  It is this property that will allow us to expand this loop integral in a series of operators with derivatives of the background field $\phi(t)$ accompanied by appropriate powers of $1/M$.  In the low energy limit, where $\phi(t)$ changes little during an interval $1/M$, operators with more derivatives will correspondingly be more suppressed. 

As a first step in this expansion, we integrate by parts with respect to the $t'$ dependence inside the loop-momentum integral.  This step simultaneously increases the convergence of the integral, by adding another power of $\omega_k$ in the denominator, while moving a derivative to the $\phi^2(t')$ factor.  The formal structure of this loop correction is exactly what we met before in analyzing the loop corrections from the $\lambda\Phi^4$ self-coupling.  Once again we define a kernel function, 
\begin{equation}
K(t-t') \equiv \int {d^3\vec k\over (2\pi)^3}\, 
{\cos\bigl[ 2\omega_k(t-t')\bigr]\over \omega_k^3} , 
\end{equation}
and integrate the loop correction to the $\phi$ equation by parts to generate three terms, 
\begin{eqnarray}
&&\!\!\!\!\!\!\!\!\!\!\!\!\!
{1\over 2} g^2 \phi(t) \int_{t_0}^t dt'\, \phi^2(t') 
\int {d^3\vec k\over (2\pi)^3}\, {\sin\bigl[ 2\omega_k(t-t')\bigr]\over \omega_k^2} 
= {1\over 4} g^2 \phi(t) \int_{t_0}^t dt'\, \phi^2(t') {d\over dt'}K(t-t')
\nonumber \\
&=&
{1\over 4} g^2 \phi^3(t) K(0) - {1\over 4} g^2 \phi(t)\phi^2(t_0) K(t-t_0)
- {1\over 4} g^2 \phi(t) \int_{t_0}^t dt'\, {d\phi^2(t')\over dt'} K(t-t') . 
\end{eqnarray}
The first term is another logarithmically divergent correction to the quartic self-interaction of the light field which can be renormalized by defining a physical coupling, 
$$
\lambda_{\rm phys} = \lambda + {\textstyle{3\over 2}} g^2 K(0) + \cdots .
$$

The second term is a mass term -- $\phi^2(t_0)$ is just a constant---but it is one with a time-dependent mass since the kernel $K(t-t_0)$ depends explicitly on the time.  We can evaluate this time-dependence exactly, once we have integrated over the loop-momentum, and express the result as a Meijer $G$-function, 
\begin{equation}
K(t-t_0) = {M(t-t_0)\over 8\pi} G^{20}_{13}\bigl( 
M^2(t-t_0)^2 \big|^{\ \ 1}_{-{1\over 2},-{1\over 2},0} \bigr) .
\end{equation}
What is relevant for the effective theory of the light field, however, is only what happens over intervals that are long compared with the natural scale of the heavy theory, which in this case means $M(t-t_0)\gg 1$.  In this limit, the leading time-dependence of the kernel can be extracted through a saddle-point approximation of the momentum integral, 
\begin{equation}
\label{eq:saddlept}
K(t-t_0) = {1\over 2\sqrt{e\pi^3}} 
{\cos\bigl[ 2M(t-t_0) + {3\pi\over 4} \bigr]\over [2M(t-t_0)]^{3/2}} 
+ \cdots .
\end{equation}

Although we have found that the two boundary terms from the integration by parts produce two corrections to the equation of motion---one local and familiar and the other nonlocal and a little less so---there still remains another unevaluated time integral.  It might seem that we have merely postponed our difficulties by pushing them into this term.  Does it also contain any further nonlocal dependence on $t-t_0$ and if so, what is it?  Does it decay and is its decay rate different from what we have already seen?

For the one-loop correction of the previous section, where the loop contained a light field, there was little to be gained by further integrating the kernel by parts, once we extracted its logarithmically divergent piece.  But when the loop field is heavy and the external legs are light, then each further integration by parts moves a time derivative from the part of the graph depending on the heavy field, which scales as ${d\over dt'}\sim M$, to the light part, which scales as ${d\over dt'}\sim m$.  Iterating this process infinitely produces a series of contributions to the equation of motion which are further and further suppressed in the low energy effective theory.

Let us define the kernels that occur in each of the successive integrations by parts to be 
\begin{eqnarray}
K^+_{3+2n}(t-t') &=& \int {d^3\vec k\over (2\pi)^3}\, 
{\cos[2\omega_k(t-t')]\over\omega_k^{3+2n}} 
\nonumber \\
K^-_{2+2n}(t-t') &=& \int {d^3\vec k\over (2\pi)^3}\, 
{\sin[2\omega_k(t-t')]\over\omega_k^{2+2n}} . 
\label{eq:nkernels}
\end{eqnarray}
These function are even ($+$) or odd ($-$) in their arguments under $t-t'\to t'-t$.  The original kernel that we introduced before is thus $K(t-t')=K^+_3(t-t')$.  Each kernel is related to the one just above or just below it by a very simple recursion relation, 
\begin{eqnarray}
K^-_{2n}(t-t') = {1\over 2} {d\over dt'} K^+_{2n+1}(t-t') 
\nonumber \\
K^+_{2n+1}(t-t') = -{1\over 2} {d\over dt'} K^-_{2n+2}(t-t') . 
\end{eqnarray}

Using these recursion relations, we integrate the one-loop correction by parts an infinite number of times to generate the following series
\begin{eqnarray}
&&\!\!\!\!\!\!\!\!\!\!\!\!
\int_{t_0}^t dt'\, \phi^2(t') 
\int {d^3\vec k\over (2\pi)^3}\, {\sin\bigl[ 2\omega_k(t-t')\bigr]\over \omega_k^2} 
= \int_{t_0}^t dt'\, \phi^2(t') K_2^-(t-t') 
\nonumber \\
&=& 
\sum_{n=0}^\infty {(-1)^n\over 2^{2n+1}} K^+_{3+2n}(0) 
{d^{2n}\phi^2(t)\over dt^{2n}}
+ \sum_{n=1}^\infty {(-1)^{n+1}\over 2^{2n}} K^-_{2+2n}(0) 
{d^{2n-1}\phi^2(t)\over dt^{2n-1}} 
\nonumber \\
&&
- \sum_{n=0}^\infty {(-1)^n\over 2^{2n+1}} K^+_{3+2n}(t-t_0) 
{d^{2n}\phi^2(t') \over dt^{\prime 2n}} \biggr|_{t'=t_0} 
- \sum_{n=1}^\infty {(-1)^{n+1}\over 2^{2n}} K^-_{2+2n}(t-t_0) 
{d^{2n-1}\phi^2(t')\over dt^{\prime 2n-1}} \biggr|_{t'=t_0}
\nonumber \\
&&
- \lim_{N\to\infty} \biggl[ 
{(-1)^{N+1}\over 2^{2N}} \int_{t_0}^t dt'\, K^-_{2+2N}(t-t') 
{d^{2N}\phi^2(t')\over dt^{\prime 2N}} 
\biggr] .  
\end{eqnarray}
Just as with the first step of this process, each subsequent integration produces two more boundary terms, one at $t'=t$ and another at $t'=t_0$.  Each of the former corresponds to a local term in the equation of motion for the light field, since the only time-dependence is in the field itself, $\phi(t)$.  All the latter however are {\it nonlocal\/} since the kernels also depend on the time that has elapsed since the initial time.  There still remains an unevaluated $dt'$ integral---the last line of this equation---but at low energies its contribution is infinitesimally small.  To understand this property, notice that the kernel within it scales as $K^-_{2+2N}\sim M^{-2N+1}$.  It also contains a $2N^{\rm th}$ time-derivative of the light field.   If the typical variation of the light field $\phi(t)$ is more generally equal to some energy scale $E$, then the integrand of this term scales as 
\begin{equation}
K^-_{2+2N}(t-t') {d^{2N}\phi^2(t')\over dt^{\prime 2N}}
\sim \biggl( {E\over M}\biggr)^{2N} M\phi^2(t') . 
\end{equation}
The effective theory is applicable for the energy regime where $E\ll M$; so as $N\to\infty$, this integrand, and the corresponding integral, essentially vanishes.

Before matching to the low energy effective theory, let us first summarize what we have found for the one-point function of the light field.  The structure that is valid for the order $g^2$ corrections is 
\begin{equation}
\langle 0(t_1)|\varphi(t_1,\vec x)|0(t_1)\rangle 
= - \int_{t_0}^{t_1}dt\, {\sin[m(t_1-t)]\over m} \Bigl\{ \cdots \Bigr\} = 0 , 
\end{equation}
where the ellipses are the loop-corrected equations of motion for the background field,
\begin{eqnarray}
\Bigl\{ \cdots \Bigr\} = 0 &=& 
\ddot\phi(t) + m^2_{\rm phys}\, \phi(t) + {1\over 6} \lambda\, \phi^3(t) 
- {1\over 4} g^2 K_3^+(0)\, \phi^3(t) 
\nonumber \\
&&
-\,\, {g^2\over 16\pi^2}\, \phi(t) \sum_{n=1}^\infty {(-1)^n\over 2^n} 
{(n-1)!\over (2n+1)!!} {1\over M^{2n}} {d^{2n}\phi^2(t)\over dt^{2n}} 
\nonumber \\
&&
+\,\, {1\over 2} g^2\, \phi(t) \sum_{n=0}^\infty {(-1)^n\over 2^{2n+1}} 
K_{3+2n}^+(t-t_0) {d^{2n}\phi^2(t')\over dt^{\prime 2n}}\Bigr|_{t'=t_0} 
\nonumber \\
&&
+\,\, {1\over 2} g^2\, \phi(t) \sum_{n=0}^\infty {(-1)^{n+1}\over 2^{2n}} 
K_{2+2n}^-(t-t_0) {d^{2n-1}\phi^2(t')\over dt^{\prime 2n-1}}\Bigr|_{t'=t_0} 
+ {\cal O}(g^3) . \ \ 
\end{eqnarray}
In this expression we have evaluated the even kernels at zero, $K_{3+2n}^+(0)$, and used the fact that the odd kernels vanish there, $K_{2+2n}^+(0)=0$, which is why only even powers of time-derivatives occur in the terms that depend only on $t$ and not $t_0$.  

What is the effective action that reproduces this expectation value?  We shall answer this question in two parts, first showing that the terms that do not depend at all on the initial time (the first two lines of this equation) are generated by a standard, local effective action as would be familiar in a setting appropriate for an $S$-matrix treatment, and then discussing the origin of the terms that depend on the initial time (the second two lines), showing that they can be produced by nonlocal, transient operators.

\subsubsection{Matching the local operators}

We should like to find the set of operators that generates exactly the same terms as occur in the one-point function that are local in time.  If we had written down all of the local, Poincar\'e-invariant operators that are consistent with the additional symmetries of the full theory, in this case its invariance under $\Phi\to -\Phi$, expanded them about the background, $\Phi(t,\vec x)=\phi(t)+\varphi(t,\vec x)$, evaluated  the resulting form of the one-point function to leading order, 
\begin{equation}
\langle 0(t_1)|\varphi(t_1,\vec x)|0(t_1)\rangle 
= \langle 0(t_0)| T_c \bigl( \varphi^+(t_1,\vec x) 
e^{-i\int_{t_0}^{t_1}dt\, [H_{I,{\rm eff}}^+(t) - H_{I,{\rm eff}}^-(t)] } \bigr) |0(t_0)\rangle , 
\end{equation}
and finally matched the coefficients of analogous terms, then we should have arrived at the following local, effective action,
\begin{eqnarray}
S_{\rm eff}^{\rm local}[\Phi] &=& 
\int_{t_0}^{t_1} dt \int d^3\vec x\, \biggl\{ 
{1\over 2} \partial_\mu\Phi\partial^\mu\Phi - {1\over 2} m^2_{\rm phys} \Phi^2 
- {1\over 24} \lambda_{\rm phys} \Phi^4 \\
&&\qquad 
+\,\, {g^2\over 64\pi^2}\, \sum_{n=1}^\infty {1\over 2^n} 
{(n-1)!\over (2n+1)!!} {1\over M^{2n}} 
\bigl( \partial_{\mu_1}\cdots\partial_{\mu_n} \Phi^2 \bigr) 
\bigl( \partial^{\mu_1}\cdots\partial^{\mu_n} \Phi^2 \bigr) +{\cal O}(g^3) \biggr\} \nonumber , 
\end{eqnarray}
where as before we have included the logarithmically divergent correction within the renormalized quartic coupling, $\lambda_{\rm phys} = \lambda + {3\over 2}g^2 K_3^+(0)$.

The original graph that we have been considering had only four external legs, three of which were for the background.  We have therefore found the leading forms, at $g^2$, for all of the quartic operators of the field.  Higher powers of the field will also be needed in the effective theory, such as $\Phi^6$, but only at higher orders in $g$.

\subsubsection{Matching the nonlocal, transient operators\/}

In writing just the Poincar\'e-invariant operators we have not captured all of the terms in the one-point function.  This in itself should not be surprising, since we have not yet entirely followed the symmetry prescription to write down an action, which is to include all the operators that are consistent with the symmetries of the theory.  While the Lagrange density of the full theory transformed as a scalar under the Poincar\'e transformations, the action was nonetheless not Poincar\'e-invariant---the seemingly minor choice to have the system start at a particular time explicitly breaks the time-translation symmetry.  Therefore, unless we are in a simple eigenstate, we must in principle include operators that are also explicitly time-dependent.  These operators are nonlocal in time and are a consequence of having heavy or high energy modes included in our initial state, which spoil the usual construction of a local effective field theory.

Were this time-dependence completely general, then it would be more or less impossible to guess the low energy theory without prior knowledge of the full theory.  However, the situation here is not quite so bleak and unconstrained.  At times much later than the initial time, the time-translation symmetry is partially restored, just so long as we do not consider translations larger than the size of the time that has elapsed since the initial time, $t-t_0$.  So we expect that as $t-t_0\to\infty$, any time-dependent effects should grow increasingly negligible.  As an effective theory, it is not meant to be applied at very short intervals, where by `short' we mean on the scale of the dynamics of the heavy field, $M(t-t_0)\le 1$.  There is thus the possibility for lingering transient effects in the intermediate regime, $1\ll M(t-t_0) \ll \infty$, as the system is settling down.  What we shall find is that the time-dependence of these operators in the effective theory is of a universal form.  Its form is set by the very specific behavior inherited from the heavy loop and the fact that the heavy field is in its free vacuum state.

The nonlocality of the corrections to the one-point function occurs in two places:  in part it is present in the fact that the field is evaluated both at $t$ and at $t_0$, but it also appears in the time-dependence of the kernels.  The leading nonlocal terms, for example, in the equations of motion are 
\begin{equation}\textstyle
- {1\over 4} g^2\, K_3^+(t-t_0) \phi^2(t_0)\phi(t) 
- {1\over 4} g^2\, K_4^-(t-t_0) \dot\phi(t_0)\phi(t_0)\phi(t) 
+ \cdots 
\end{equation}
From the perspective of the classical field, both of these terms look like time-dependent masses, since they are linear in $\phi(t)$. However, the fact that there are fields evaluated at the initial times is important too---it is meant to guide our guess for the form of the transient operators in the effective theory.  Let us consider the time-dependence in the kernels first.

Once again the kernels can be evaluated at an arbitrary time and expressed in Meijer $G$-functions,
\begin{eqnarray}
K^+_{3+2n}(t-t_0) &=& 
{1\over 8\pi} {[M(t-t_0)]^{2n+1}\over M^{2n}} 
G^{20}_{13}\bigl( M^2(t-t_0)^2 \big|^1_{-{1\over 2}, -n-{1\over 2}, -n} \bigr)
\nonumber \\
K^-_{2+2n}(t-t_0) &=& 
{1\over 8\pi} {[M(t-t_0)]^{2n}\over M^{2n-1}} 
G^{20}_{13}\bigl( M^2(t-t_0)^2 \big|^1_{-{1\over 2}, -n+{1\over 2}, -n} \bigr) . 
\end{eqnarray}
But as was just mentioned, what is applicable for the effective theory is what happens in the limit $M(t-t_0)\gg 1$ where we can perform a saddle-point evaluation of the spatial momentum integrals, yielding a universal form for the kernels,
\begin{eqnarray}
K^+_{3+2n}(t-t_0) &=& 
{1\over 2\sqrt{e\pi^3}} {1\over M^{2n}} 
{\cos\bigl[ 2M(t-t_0)+{3\pi\over 4}\bigr]\over [2M(t-t_0)]^{3/2}} 
+ {\cal O}\bigl( [M(t-t_0)]^{-5/2} \bigr)
\nonumber \\
K^-_{2+2n}(t-t_0) &=& 
{1\over 2\sqrt{e\pi^3}} {1\over M^{2n-1}} 
{\sin\bigl[ 2M(t-t_0)+{3\pi\over 4}\bigr]\over [2M(t-t_0)]^{3/2}} 
+ {\cal O}\bigl( [M(t-t_0)]^{-5/2} \bigr) . 
\end{eqnarray}
There is only a relative displacement by ${\pi\over 2}$ between the phases of the even and odd kernels.  The leading parts of each of these kernels decay at precisely the same rate, $(t-t_0)^{-3/2}$.  

The origin of this universal behavior is easy to understand, and it is worth explaining this origin to gain a little insight into why we can still have control of the structures in the effective theory in time-dependent settings.  Being an infrared effect---we are looking in the large $t-t_0$ limit---most of the contribution to the momentum integral comes from the small $k$ region.  In this region the integrand does not depend sensitively on the order of the kernel, $n$.  More explicitly, after we have integrated over the angular directions, the even kernel for example can be written as 
\begin{equation}
K_{3+2n}^+(t-t_0) = {1\over 2\pi^2} \int_0^\infty dk\, 
{k^2\over (M^2+k^2)^{n+{3\over 2}}} \cos\bigl[ 2\sqrt{M^2+k^2}(t-t_0)\bigr] ,
\end{equation}
or, in terms of the dimensionless variable $\kappa = k/M$, as 
\begin{equation}
K_{3+2n}^+(t-t_0) = {1\over 2\pi^2} {1\over M^{2n}} 
\int_0^\infty d\kappa\, {\kappa^2\over (1+\kappa^2)^{n+{3\over 2}}} 
\cos\bigl[ 2M(t-t_0)\,\sqrt{1+\kappa^2}\bigr] . 
\end{equation}
Notice that the prefactor shows how it scales in powers of the heavy mass.  For large $M(t-t_0)$ there is a strong oscillatory suppression from the cosine when $\kappa$ is large.  For small $\kappa$, the order of the kernel is a higher-order effect, $(1+\kappa^2)^{n+{3\over 2}}\approx 1$, and it is the behavior of the integration measure $\kappa^2\, d\kappa$ that is most decisive, determining the scaling to be $(t-t_0)^{-3/2}$.  This crude reasoning can be verified by a more careful saddle-point evaluation of this integral.  Had the kernel been defined in $d$ spatial dimensions, for example, then the asymptotic time-dependence would have been a sinusoid multiplied by $(t-t_0)^{-d/2}$ instead.

So the time-dependence of all of the nonlocal corrections to the equations at this order has a common structure.  Once we are within the regime applicable to the effective theory, $M(t-t_0)\gg 1$ the aforementioned nonlocal corrections become,
\begin{eqnarray}
&&
- {g^2\over 8\sqrt{e\pi^3}} 
{\cos\bigl[ 2M(t-t_0)+{3\pi\over 4}\bigr]\over [2M(t-t_0)]^{3/2}}
\phi(t) \sum_{n=0}^\infty \biggl( -{1\over 4}\biggr)^n {1\over M^{2n}} 
{d^{2n}\phi^2(t') \over dt^{\prime 2n}} \biggr|_{t'=t_0} 
\\
&&
- {g^2\over 8\sqrt{e\pi^3}} 
{\sin\bigl[ 2M(t-t_0)+{3\pi\over 4}\bigr]\over [2M(t-t_0)]^{3/2}}
\phi(t) \sum_{n=0}^\infty {1\over 2} \biggl( -{1\over 4}\biggr)^n {1\over M^{2n+1}} 
{d^{2n+1}\phi^2(t')\over dt^{\prime 2n+1}} \biggr|_{t'=t_0} 
+ \cdots ,
\nonumber 
\end{eqnarray}
where we have written only the leading term in powers of $M(t-t_0)$ for each order in the ${1\over M}{d\over dt}$ expansion.  At the level of the equations of motion for the background, these terms would all appear to be essentially the parts of a single time-dependent mass term, since $\phi(t_0)$ and its derivatives are all constant and all the terms are linear in $\phi(t)$.  However, while in the case of the background $\phi^2(t_0)$ and its derivatives are just constants, they are really meant to act as substitutes in the place of real propagating fields.  To be able to match higher order $N$-point functions of $\varphi(t,\vec x)$, we shall require operators that are explicitly nonlocal in time in the effective theory.  For example, the leading nonlocal operator in the effective theory is 
\begin{equation}
S_{\rm eff}^{\rm nonlocal}[\Phi] = \int_{t_0}^{t_1}dt \int d^3\vec x\, 
\Bigl\{ - {\textstyle{1\over 2}} \mu^2(t-t_0)\, \Phi^2(t_0)\Phi^2(t) 
+ \cdots \Bigr\} .
\end{equation}
Upon expanding this operator about the background and evaluating its contribution to the one-point function in the effective theory, we find
\begin{equation}
\langle 0(t_1)|\varphi(t_1,\vec x)|0(t_1)\rangle 
= - \int_{t_0}^{t_1}dt\, {\sin[m(t_1-t)]\over m} \Bigl\{ \mu^2(t-t_0)\, \phi^2(t_0)\phi(t) + \cdots \Bigr\} .
\end{equation}
Matching with the appropriate term in the one-point function calculated in the full theory, we see that 
\begin{equation}
\mu^2(t-t_0) = - {g^2\over 8\sqrt{e\pi^3}} 
{\cos\bigl[ 2M(t-t_0)+{3\pi\over 4}\bigr]\over [2M(t-t_0)]^{3/2}}
+ \cdots 
\end{equation}
where we have again already written its form in the regime relevant for the effective theory.  So we see that there is a short-lived operator which has resulted from the breaking of Poincar\'e invariance by the presence of the initial time.

Just as was true of the order $g^2$ operators, this operator is but the first instance of a tower of very similar operators suppressed by powers of $\partial_\mu/M$.  Unlike the Poincar\'e invariant operators, though, this set does include terms with odd numbers of time derivatives.  The normal to the initial-time hypersurface, $n^\mu$, effectively provides another vector which can be contracted with a derivative, $n^\mu\partial_\mu$.

The effective action to order $g^2$ is given by the sum of its local and nonlocal parts,
$$
S_{\rm eff}[\Phi] 
= S_{\rm eff}^{\rm local}[\Phi] + S_{\rm eff}^{\rm nonlocal}[\Phi] .
$$
The local part of this action is exactly what would have been anticipated by a standard $S$-matrix treatment, but it would have overlooked the nonlocal terms since it usually assumes that the action remains fully Poincar\'e invariant, especially over short intervals.  For an $S$-matrix calculation, following the standard effective theory prescription, there would be no reason to include operators in the low energy action that possess less symmetry than the full action.

As mentioned before, the explicit time-dependence in this example occurs only rather subtly, in the limits of the time integration; the Lagrange density is itself still a scalar under Poincar\'e transformations.  We can make this time-dependence more explicit by rewriting the interaction as 
$$
\int_{t_0}^\infty dt\int d^3\vec x\, \bigl\{ 
\cdots + {\textstyle{1\over 2}}g\chi^2\Phi^2 + \cdots \bigr\} 
= \int_{-\infty}^\infty dt\int d^3\vec x\, \bigl\{ 
\cdots + {\textstyle{1\over 2}}g\Theta(t-t_0)\, \chi^2\Phi^2 + \cdots \bigr\} . 
$$
From this form it should be clear that we are regarding the $t_0$ as being associated with a genuine physical change in the theory and one that happens infinitely quickly, rather than just providing a convenient starting point.  In the latter case, the $t_0$ dependence would not carry any physical information and no physical measurement should depend on it; the Poincar\'e invariance would be unbroken and the nonlocal part of the effective theory would not be needed.  But this case corresponds to a different physical situation than the one that we have examined here.

The fact that the time-dependence is fast on the scale of the heavy field is important for the existence of the nonlocal operators.  For example, suppose that we had instead turned the coupling, $g(t)$, on more gradually between $t_{\rm off}$ and $t_{\rm on}$ where
$$
g(t) = \bigl\{ g_0\ \hbox{if $t\ge t_{\rm on}$}, \quad
0\ \hbox{if $t\le t_{\rm off}$} \bigr\} . 
$$
In this case, the one-point function, evaluated to order $g^2$, would produce an equation for the background of  
\begin{equation}
\ddot\phi(t) + m^2\phi(t) + {\textstyle{1\over 6}}\lambda\phi^3(t) 
+ {\textstyle{1\over 2}} g_0\phi(t) 
\int_{t_0}^t dt'\, g(t')\phi^2(t') K_2^-(t-t') 
+ \cdots 
\end{equation}
where we have assumed that $t>t_{\rm on}$ and $t_0<t_{\rm off}$.  Concentrating on the integral in the last term, if we integrate the kernel by parts an infinite number of times, we find
\begin{eqnarray}
&&\!\!\!\!\!\!\!\!\!\!\!\!\!\!\!\!\!\!\!
{1\over 2} g_0\phi(t) \int_{-\infty}^t dt'\, g(t')\phi^2(t') K_2^-(t-t') 
\nonumber \\
&=& 
{1\over 4} g_0^2K_3^+(0)\phi^3(t) 
+ {g^2_0\over 16\pi^2} \phi(t) \sum_{n=1}^\infty {(-1)^n\over 2^n} 
{(n-1)!\over (2n+1)!!} {1\over M^{2n}} {d^{2n}\phi^2(t)\over dt^{2n}}
\nonumber \\
&&
+ {1\over 2} g_0\phi(t) \lim_{N\to\infty}\biggl[ {(-1)^N\over 2^{2N}} \int_{t_0}^t dt'\, K_{2+2N}^-(t-t') 
{d^{2N}\over dt^{\prime 2N}} \Bigl[ g(t') \phi^2(t') \Bigr] \biggr] .
\end{eqnarray}
We have not included any of the terms at the initial boundary at $t=t_0$ since we are assuming that $g(t)$ and all of its derivatives vanish there.  Within the integrand of the final term the kernel scales as $M^{-2N+1}$.  As long as the derivatives of $g(t)$ are sufficiently smaller than $M$, then this term vanishes in the $N\to\infty$ limit.  Having $g(t)$ vary on the same time-scales as the background $\phi(t)$ is certainly sufficient.  So if we turn on the interactions slowly compared with the scale of the heavy field, we are left with only the local terms in the effective action at this order.

Another way of looking at the origin of these nonlocal structures is that in a Poincar\'e invariant theory there is only the scale $M$ from which to build up an effective theory, using dimensionless objects such as $\Phi/M$ or $\partial_\mu/M$. In a theory with an initial time, we have an additional scale $t-t_0$ and therefore there is an additional dimensionless parameter $M(t-t_0)$.  We also have a new vector $n_\mu$ normal to the initial-time hypersurface.  So the possibility of a dependence of the parameters on $(t-t_0)$ is not in itself surprising.  But what we have found is that this time-dependence has a simple, universal form:  a sinusoid whose amplitude decays as $[M(t-t_0)]^{-3/2}$.  These nonlocal terms therefore have two interesting properties.  They do decay, so that with sufficient time the theory returns to a Poincar\'e invariant form.  But their decay is also rather slow, and nonanalytic, so some of them might last long enough to generate some interesting transient effects depending upon the system and observable quantities we are analyzing.  At higher orders in the coupling $g$, we shall meet further nonlocal corrections too.  This subject we investigate next.

\subsection{The one-point function at higher orders}

Although we shall not examine the higher-order one-loop corrections with quite the same thoroughness with which we have just treated the leading corrections, it is nonetheless worth looking a little at what new arises when we add further external legs for the background.  We shall see in part that the same pattern and structures of the $g^2$ graph are reproduced at higher orders, but we shall also meet another class of nonlocal corrections, which might not have been anticipated from the lower order corrections.

\begin{figure}[htbp]
\begin{center}
\includegraphics[scale=1]{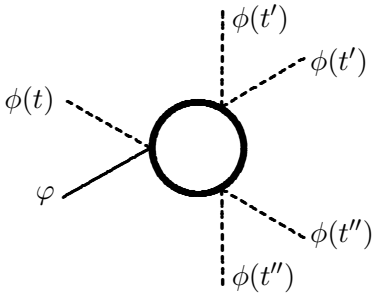}
\caption{The one-loop, order $g^3$ contribution to the one-point function.}
\label{loopthree}
\end{center}
\end{figure}

At the next order, at order $g^3$, we encounter a one-loop graph with five external $\phi$-legs, as shown in figure  \ref{loopthree}.
Once we have amputated the propagating $\varphi$-leg, we find the following contribution to the equation of motion for the background $\phi(t)$ from this graph, 
\begin{eqnarray}
&&\!\!\!\!\!\!\!\!\!\!\!\!\!\!\!\!\!\!\!\!\!
{1\over 2} g^3 \phi(t) \int_{t_0}^t dt'\, \phi^2(t') 
\int_{t_0}^{t'} dt^{\prime\prime}\, \phi^2(t^{\prime\prime}) K_3^+(t'-t^{\prime\prime})
\nonumber \\
&&
- {1\over 2} g^3 \phi(t) \int_{t_0}^t dt'\, \phi^2(t') 
\int_{t_0}^{t'} dt^{\prime\prime}\, \phi^2(t^{\prime\prime}) K_3^+(t-t^{\prime\prime}) .  
\end{eqnarray}
The time-dependence of the kernels is slightly different, and we now have two intermediate times over which to integrate. However,  because no spatial momentum flows into the diagram, the kernel functions are exactly the same spatial loop-momentum integrals that we have already defined.

The integral on the first line can be converted into an infinite series of terms just as before.  Some will be entirely local, where the time-dependence only occurs in the field and its derivatives.  Some will be partially nonlocal, where the time-dependence is still only in the field and its derivatives, but where some of the fields are also evaluated at the initial time.  And lastly, some of the terms will be more strongly nonlocal in the sense that they will be augmented by a time-dependence from the $(t-t_0)$-dependent kernel.  All of these terms can be grouped into a series of double sums, since we have two time-integrals to expand, but otherwise all is familiar from the order $g^2$ analysis.

The term in the second line, however, generates a qualitatively different correction, which is more difficult to reproduce from a simple effective theory.  Suppose that we attempt to perform a derivative expansion of this term.  One of the surface terms generated by the very first of the partial integrations is 
\begin{eqnarray} 
&&\!\!\!\!\!\!\!\!\!\!\!\!\!\!\!\!\!\!\!\!\!\!\!\!\!\!
- {1\over 2} g^3 \phi(t) \int_{t_0}^t dt'\, \phi^2(t') 
\int_{t_0}^{t'} dt^{\prime\prime}\, \phi^2(t^{\prime\prime})  K_3^+(t-t^{\prime\prime}) 
\nonumber \\
&=&
- {1\over 4} g^3\phi^2(t_0) K_4^-(t-t_0)\phi(t) \int_{t_0}^t dt'\, \phi^2(t') 
+ \cdots .  
\end{eqnarray}
When we arrive at this term, it is not possible to proceed any further in its expansion.  In addition to its ${1\over 4} g^3\phi^2(t_0) K_4^-(t-t_0) \phi(t)$ prefactor, which looks like another transient nonlocal correction, we have an integral of $\phi^2(t')$.  Its integrand is always positive, and the range of the integral grows with the time that has elapsed since the initial time.

Let us see how this term fits amongst the other order $g^3$ corrections.  Including the order $g^0$ terms for reference, the leading $g^3$ corrections to the $\phi(t)$ equation of motion are the following, 
\begin{eqnarray}
&&\!\!\!\!\!\!\!\!\!\!\!\!\!\!\!\!\!\!\!\!\!
\ddot\phi(t) + m^2 \phi(t) + {1\over 6} \lambda \phi^3(t) + 
\hbox{order $g^2$ terms}
\nonumber \\
&&
+ {1\over 48\pi^2} {g^3\over M^2} \phi^5(t)
+ {1\over 48\pi^2} {g^3\over M^2} \phi^4(t_0)\phi(t)
\nonumber \\
&&
- {1\over 8} g^3 K_5^+(t-t_0) \phi^2(t_0) \phi^3(t) 
- {1\over 8} g^3 K_5^+(t-t_0) \phi^4(t_0) \phi(t) 
\nonumber \\
&&
- {1\over 4} g^3 \phi^2(t_0)  K_4^-(t-t_0) \phi(t) \int_{t_0}^t dt'\, \phi^2(t') 
+ \cdots = 0 . 
\end{eqnarray}
We have left out the higher order corrections in the derivative expansion.  The second line shows the two leading corrections which are local at the level of the equation of motion for $\phi(t)$.  The first corresponds to the correction produced by a $\Phi^6$ operator in the effective theory.  The second resembles a mass term, since its only time-dependence is in $\phi(t)$, but at a deeper level it has some nonlocality too, since the field is evaluated at the initial time in the $\phi^4(t_0)$ factor.

Neglecting this term, since at the level of the equation of motion it acts only as a small additional correction to the local mass term, and the integral term on the last line, we can ask what sorts of operators would produce this same set of corrections if we evaluated the one-point function using a low energy theory.  This reasoning would lead to the following set of operators,
\begin{eqnarray}
S_{\rm eff}^{{\cal O}(g^3)}[\Phi] 
&=& \int_{t_0}^\infty\ dt d^3\vec x\, \biggl\{
- {1\over 18}{1\over 16\pi^2} {g^3\over M^2} \Phi^6(t,\vec x) 
+ {1\over 6!} {g^3\over M^2}c_3(t-t_0)\Phi^2(t_0,\vec x)\Phi^4(t,\vec x)
\nonumber \\
&&\qquad\qquad\qquad\!\!
+ {1\over 6!} {g^3\over M^2}\tilde c_3(t-t_0)\Phi^4(t_0,\vec x)\Phi^2(t,\vec x)
+ \cdots \biggr\} . 
\end{eqnarray}
The time-dependence of the coefficients in the nonlocal terms is precisely that inherited from the kernels.  At intermediate times, 
$1\ll M(t-t_0)\ll \infty$,  
\begin{eqnarray}
c_3(t-t_0) 
&=&
{45\over 2} M^2 K_5^+(t-t_0) 
= {45\over 4\sqrt{e\pi^3}}  
{\cos\bigl[ 2M(t-t_0) + {3\pi\over 4}\bigr]\over [2M(t-t_0)]^{3/2}} 
+ \cdots 
\nonumber \\
\tilde c_3(t-t_0) 
&=&
45 M^2 K_5^+(t-t_0) 
= {45\over 2\sqrt{e\pi^3}}  
{\cos\bigl[ 2M(t-t_0) + {3\pi\over 4}\bigr]\over [2M(t-t_0)]^{3/2}} 
+ \cdots . 
\end{eqnarray}
The part of the second of the nonlocal terms that depends on $t$, as opposed to $t_0$, is $\Phi^2(t,\vec x)$, just like the analogous term that appeared at order $g^2$.  But now we also have a further nonlocal, nonderivative term which is quartic in its time-dependent part.  In general, the highest power of the time-dependent parts of the nonlocal operators that appear at a given order in $g$ is two powers lower than for the corresponding local operator.  

Returning to the integral term, notice that it does not have an immediate interpretation as an operator in the effective theory, even allowing for some time-dependence in the couplings.  This term is not unique. Rather, it is only the first instance of a family of corrections of an analogous form, appearing at higher orders, with successively more powers of the field and more nested time integrals.  For example, at order $g^4$, we encounter a correction to the background equation of motion with two nested time integrals,
\begin{equation}
{1\over 4} g^4 \phi(t)\phi^2(t_0) K_5^+(t-t_0) 
\int_{t_0}^t dt'\, \phi^2(t') 
\int_{t_0}^{t'} dt^{\prime\prime}\, \phi^2(t^{\prime\prime}) 
\end{equation}
together with other corrections containing only one time integral but more powers of the field,
\begin{eqnarray}
&&\!\!\!\!\!\!\!\!\!\!\!\!\!\!\!
-\,\, {1\over 16} g^4 \phi(t)\phi^2(t_0) K_6^-(t-t_0) 
\int_{t_0}^t dt'\, \phi^4(t') 
\nonumber \\
&&
-\,\, {1\over 8} g^4 \phi^2(t_0) K_6^-(t-t_0) 
\bigl\{ \phi^3(t) + \phi(t)\phi^2(t_0) \bigr\}
\int_{t_0}^t dt'\, \phi^2(t') 
+ \cdots . 
\end{eqnarray}

This glimpse at a few of the corrections that appear at higher orders in the coupling suggests the richness of the structures that are possible in a time-dependent setting.  In the next section we shall examine them from a more general perspective, delineating what is required of the effective theory by each class of corrections to the one-point function found here.

\section{General classes of structures in the one-point function}
\label{sec:systematics} 

At second order in the coupling between the light and heavy parts of the theory, the corrections to the equation of motion can be characterized as being either local or nonlocal in time.  All of the nonlocal corrections were seen to decay at a uniform rate of $[M(t-t_0)]^{-3/2}$.  When one proceeds to the higher order corrections, one sees that this pattern is only in part reproduced.  There are again local corrections to the equation of motion, though unlike the order $g^2$ case, some of these terms also begin to include fields evaluated at the initial time too, suggesting that they come from partially nonlocal terms in the action.  There are also nonlocal operators that, as before, are nonlocal both in the time-dependence of the fields and in their coefficients, which are suppressed by more powers of the heavy mass as well as the same $[M(t-t_0)]^{-3/2}$ decay as at order $g^2$.  So at any time, and given a finite experimental precision, only a finite number of these terms will have a detectable influence.  Since they are decaying, the number that have an observable effect diminishes until one is essentially left with only the need to treat the local parts of a correlation function for any practical measurement.

We have seen that there is a second type of nonlocal contribution too, which has an integral form.  They appear individually to violate our expectation that after a sufficiently long interval the theory should look approximately Poincar\'e invariant.  However, nothing should be genuinely diverging at low energies in the full theory.  Even from the perspective of having turned on the interactions suddenly, we are only adding a finite energy density to the system, though as this is being done instantaneously, the energy is being distributed through a full range of scales, including heavy ones.  It might be possible then that these terms can be summed into a function that does decay, allowing the symmetry to be restored after a suitable interval.  We shall examine the origins of both of these classes of nonlocal effects here, as well as the origin of the more ordinary local terms.

One lesson from our study of the two-vertex graph is that it is useful to separate any integrand depending on an intermediate time into a kernel function, which contains all of the dependence on the mass of the heavy field, and product of light fields; in that earlier case this product was just $\phi^2(t')$.  By integrating by parts, we can then extract powers of $1/M$ as we move higher and higher in the tower of kernels, while at the same time adding more and more derivatives to the light fields.  We thereby generate an expansion in the derivatives of the light field, where some are evaluated at $t$ (the time that appears in the field $\varphi(t,\vec x)$ whose tadpole function is being calculated) while others are evaluated at the initial time $t_0$.

\begin{figure}[htbp]
\begin{center}
\includegraphics[scale=1]{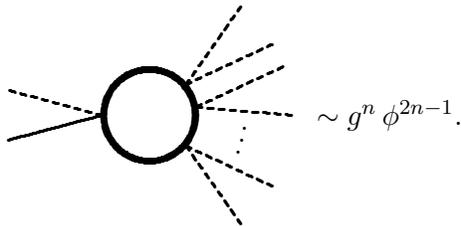}
\caption{A generic tadpole graph giving a contribution to the $\phi^{2n-1}$ term in the equation of motion.  Each vertex occurs at a different intermediate time, so $g^n\phi^{2n-1}\to g^n\phi(t)\phi^2(t_1)\cdots\phi^2(t_{n-1})$.}
\label{fig:2npt}
\end{center}
\end{figure}

Let us consider a general one-loop contribution to the tadpole function that has $n$ vertices, as shown in figure~\ref{fig:2npt}.  
As a correction to the equation of motion for the background field, it scales as $g^n\phi^{2n-1}$.  In a Poincar\'e invariant, $S$-matrix treatment, such a graph would require a local operator $c_ng^n\Phi^{2n}$, where the coefficient $c_n$ is found by matching between the full and effective theories.

In a time-dependent setting, what this graph requires of the effective theory is a bit more complicated.  The loop corresponds to an integral over the spatial momentum.  When the background field itself has no spatial dependence, then the structure of this loop integral is comparatively simple since no external momentum flows into or out of the diagram.  In contrast to this simple spatial dependence, each of the $n$ vertices occurs at a different intermediate time, $\{t,t_1,t_2,\ldots, t_{n-1}\}$.  Once we have amputated the external leg for the fluctuations, we are left with $n-1$ ordered time integrals running between $t_0$ and $t$.

For the low energy limit, we follow the same procedure that was used for the $n=2$ vertex graph, expressing the loop momentum integral through the kernels defined earlier.  The corrections to the equation of motion for the background $\phi(t)$ that emerge as we proceed to integrate each of these $n-1$ integrals by parts---as far as is possible---fall into three general classes, distinguished by the fate of the kernel.  In two of these classes we convert all of the time integrals into multiple series of fields evaluated at one or both of the boundaries.  If the kernel accompanying one of the terms has the form $K^+_{3+2n}(0)$, then that correction is a local one at the level of the equation of motion.  If instead the kernel has the form $K^+_{3+2n}(t-t_0)$ or $K^-_{2+2n}(t-t_0)$, then we have a transient, nonlocal correction, where part of the nonlocality appears in the kernel function, as before.

The third class consists of corrections where after an integration by parts the kernel moves outside all of the remaining intermediate time integrals, to become at once $K^+_{3+2n}(t-t_0)$ or $K^-_{2+2n}(t-t_0)$.  When this happens, the process of partial integrations comes to an end, if we wish still to be able to amputate the external $\varphi$-leg.  At the level of the equation of motion, while the kernel functions are still themselves decaying, they are now multiplying nested integrals over products of the light field.  As we discovered in the last section, such terms only appear at order $g^3$ and higher.  Let us describe each of these three classes in turn.

\subsection{Local corrections to equation of motion}

Unlike the two-vertex corrections that we covered in detail, in the course of expanding the time-integrals in a series of local terms, some of the terms at higher orders in the coupling will also depend on the initial value of the field or its derivatives.  For example, at order $g^3$ we saw that there is a mass term scaling as $g^3M^{-2}\phi^4(t_0)\phi(t)$.  Factors of $\phi(t_0)$ and its derivatives are just constants in the background equation of motion, but they are indicative of fields evaluated at the initial time, $\Phi(t_0,\vec x)$, which would be needed in matching for a general $N$-point function.  What characterizes these corrections is that all of the remaining time-dependence still occurs only in the field or its derivatives and not in the kernels.  Because the odd kernels vanish when their argument vanishes, the local corrections to the equation of the background will only have even numbers of derivatives acting on $\phi(t)$ and $\phi(t_0)$,
\begin{equation}
{g^n\over M^{2n-4}} \phi(t) \sum_{k=0}^\infty {1\over M^{2k}}  
\sum_{l=0}^k \sum_{m=0}^{n-1} c^{(n)}_{k,l,m} 
\biggl[ {d^{2(k-l)}\over dt^{2(k-l)}}\phi^{2n-2m-2}(t) \biggr] 
\biggl[ {d^{2l}\over dt^{\prime 2l}}\phi^{2m}(t') \biggr]_{t'=t_0} .
\end{equation}
The lone $\phi(t)$ in front corresponds to the background field attached to the external $\varphi(t,\vec x)$ leg, as is shown on the left side of the graph in figure \ref{fig:2npt}.  Note that the time derivatives in this expression are meant to be represented only suggestively---they will generally be partitioned in some way among the fields on which they are shown to be acting.

At the level of the equation of motion for the background, the terms evaluated at the initial time are already higher order corrections compared with others with the same time-dependence.  If we assume that $\dot\phi\sim E\phi$, where $E$ is some low energy scale, then a term with a $2l^{\rm th}$ derivative evaluated at the initial time will be suppressed by $g^l(E/M)^{2l}$ relative to a term already present at a lower order in $g$ among the corrections.  The leading terms for the equation of motion at a given order in $g$ then come from the $l=0$ terms in the series.

The operators of a low energy effective action that would reproduce these same $l=0$ terms, were we to have evaluated its one-point function, are 
\begin{equation}
S_{\rm eff}^{\rm local}[\Phi] 
= M^4 \int d^4x\, \biggl\{ 
\sum_{n=2}^\infty \sum_{k=0}^\infty 
c^{(n)}_k g^n \biggl[ {\square^2\over M^2} \biggr]^k 
\biggl[ {\Phi\over M} \biggr]^{2n} 
+ \cdots 
\biggl\} . 
\end{equation}
Again, the derivatives have been written only heuristically, merely counting the number present at a given order in the series of operators.  In general, they will be distributed among the fields in a particular way, or several inequivalent ways.  The exact form of the coefficients $c_k^{(n)}$ are found by matching the analogous terms of the full and effective theories.  The fact that only terms with even numbers of time-derivatives acting on $\phi(t)$ appeared in the one-point function was just what was needed to match with effects produced by Poincar\'e invariant operators in the low energy theory.

\subsection{Transient, nonlocal corrections}

The second set of corrections generated by a graph with $n$ vertices generalizes the decaying $(t-t_0)^{-3/2}$ terms.  This set has the form 
\begin{equation}
{g^n\over M^{2n-4}} \phi(t) \sum_{k=0}^\infty {1\over M^k} 
\sum_{l=0}^k \sum_{m=1}^{n-1} d^{(n)}_{k,l,m}(t-t_0) 
\biggl[ {d^{k-l}\over dt^{k-l}} \phi^{2n-2m-2}(t) \biggr] 
\biggl[ {d^l\over dt^{\prime l}}\phi^{2m}(t') \biggr]_{t'=t_0} . 
\end{equation}
Although this series of corrections superficially has much in common with the form of the local corrections, it has three differences from them, two of which are relatively minor.  First, when matching to an effective theory, the highest time-dependent power of $\Phi(t,\vec x)$ that appears at order $g^n$ in the effective action is $\Phi^{2n-2}(t,\vec x)$, and not $\Phi^{2n}(t,\vec x)$ as was the case for the local graphs.  For example, at order $g^2$ it was a nonlocal $\Phi^2(t_0,\vec x)\Phi^2(t,\vec x)$ term that appeared; at order $g^3$ we also encounter a nonlocal $\Phi^2(t_0,\vec x)\Phi^4(t,\vec x)$ operator.  The reason is that to produce a time-dependent kernel, at least one of the boundary terms from the partial integrations must be evaluated at the initial time.  Notice also that the terms can have an odd number of time derivatives, both on each of the powers of $\phi(t)$ and $\phi(t_0)$ and in total.  Again, the normal to the initial-time hypersurface effectively provides an additional vector that we can contract with a derivative.

The third difference is, of course, that the coefficients of these terms are nonlocal---the coefficients $d^{(n)}_{k,l,m}(t-t_0)$ are also time-dependent.  This dependence comes directly from the kernels; in the intermediate regime where $1\ll M(t-t_0)\ll\infty$, the coefficients of all of these nonlocal terms decay with exactly the same $(t-t_0)^{-3/2}$ behavior that we met before, 
\begin{equation}
d^{(n)}_{k,l,m}(t-t_0) \propto 
\Bigl\{ K^+_{3+2n}(t-t_0)\ \hbox{or}\ K^-_{2+2n}(t-t_0) \Bigr\}
\propto {1\over [2M(t-t_0)]^{3/2}} .
\end{equation}

The leading nonlocal operators, those without any time-derivatives, are comparatively simple to write in the low energy theory, having the form 
\begin{equation}
S_{\rm eff}^{\rm nonlocal}[\Phi] 
= M^4 \int_{t_0}^\infty dt\int d^3\vec x\, \biggl\{ 
\sum_{n=2}^\infty d^{(n)}(t) g^n {\Phi^2(t_0,\vec x)\over M^2} 
\biggl[ {\Phi^2(t,\vec x)\over M^2} \biggr]^{n-1} 
+ \hbox{derivative operators} \
\biggl\} . 
\end{equation}
The derivative operators are straightforward enough to determine in principle, though more complicated in their form, since in the process of varying the action, the standard integration by parts done to extract the linear terms in $\varphi$ can move derivatives to the time-dependent coefficients too.

In a standard effective theory with nonrenormalizable operators, the theory remains predictive since only a finite set of them have an observable effect for a given experimental precision.  For the nonlocal operators here, this condition becomes a time-dependent one.  For example, if $\dot\phi\sim E\phi$, and the size of $\phi$ is itself much smaller than $M$, then a typical nonlocal operator at order $g^n$ and with $k$ derivatives will scale as 
\begin{equation}
{g^n\over [M(t-t_0)]^{3/2}} \biggl( {E\over M} \biggr)^k 
\biggl( {|\!|\phi|\!|\over M} \biggr)^n
\end{equation}
With enough time, their influence will be too small and after that point the entire set can be ignored.  But for a sufficiently sensitive experiment, during an intermediate regime, they do need to be considered.

\subsection{Integral corrections}

So far the time dependence of the setting in the full theory has appeared in only a relatively mild form in the structures of the effective theory.  The third class of corrections to the background equation of motion exhibit a much stronger violation of the locality, in the sense that they grow with time.  Since the full theory ought not to be genuinely diverging in this limit, we suspect that these terms can be organized into a decaying function.\footnote{Another possibility is that we cannot amputate the external $\varphi$ leg for these terms and must match the unamputated one-point functions between the full and effective theories.}  A similar behavior occurs in other time-dependent systems \cite{Boyanovsky:2003ui}, where individual terms of a perturbative expansion that are growing as powers of the elapsed time can be organized into a sum where the underlying decay of these terms is made evident and the range for the applicability of the theory can be extended dramatically.

For the derivative expansion of the nested time integrals to work, the kernel must always depend on at least one of the intermediate times being integrated until they have all been exhausted.  But at order $g^3$ and above, after a single partial integration of certain kernels, the process ends at once, producing a correction of the general form,
\begin{equation}
{g^n\over M^{2n-m-4}} \tilde d^{(n)}_{[k_i],m}(t-t_0) 
\phi(t) \int_{t_0}^t dt_1\, \phi^{2k_1}(t_1) 
\int_{t_0}^{t_1} dt_2\, \phi^{2k_2}(t_2) 
\cdots 
\int_{t_0}^{t_{m-1}} dt_m\, \phi^{2k_m}(t_m) 
\end{equation}
where 
$$
k \equiv \sum_{i=1}^m k_i \le n-2 .
$$
The initial $\phi(t)$ is again just the field that occurs at the vertex with the fluctuation, $\varphi(t,\vec x)$, and the time dependence in the coefficient is inherited directly from one of the kernels, 
\begin{equation}
\tilde d^{(n)}_{k,l,m}(t-t_0) \propto 
\Bigl\{ K^+_{3+2n}(t-t_0)\ \hbox{or}\ K^-_{2+2n}(t-t_0) \Bigr\} .
\end{equation}
But while these kernels are decaying, this decay is by itself no longer sufficient to overcome the growth in the rest of the correction.  For example, the growth of the integrals that we found at order $g^4$
\begin{equation}
\int_{t_0}^t dt'\, \phi^2(t') \int_{t_0}^{t'} dt^{\prime\prime}\, \phi^2(t^{\prime\prime})
\end{equation}
is sufficient to overcome the $(t-t_0)^{3/2}$ decay of the coefficient, at least in the small coupling limit where we can use the equation of motion to solve for $\phi(t)$ perturbatively.

Although these terms are individually growing, this growth does not necessarily imply that these structures will appear in the low energy effective theory.  As an example of a way in which the sum of all such terms might conspire to rid the theory of an apparent infrared (that is, long time) divergence, let us see what their leading time-dependence should be.  The equation of motion for the background produced by the tadpole condition at zeroth order in the coupling, $\ddot\phi + m^2\phi \approx 0$, has as its solution a simple sinusoid,
$$
\phi(t) = \phi_0\cos(mt) + {\cal O}(g,\lambda) .
$$
When this solution is substituted into one of these nested integrals, we find that 
\begin{equation}
{g^k\over M^k} \int_{t_0}^t dt_1\, \phi^2(t_1) 
\int_{t_0}^{t_1} dt_2\, \phi^2(t_2) 
\cdots 
\int_{t_0}^{t_{m-1}} dt_k\, \phi^2(t_k) 
= {1\over 2^k k!} {g^k\over M^k}\phi_0^{2k} (t-t_0)^k + \cdots ,
\end{equation}
where we have not included the initial numerical prefactor or---more importantly---its sign.  Superficially this term looks disastrous if our hope is to be able to match it with a sensible operator in the low energy effective theory.  It grows much faster than the mild decay of the kernel.  However, notice that exactly the same terms occur in the expansion of a {\it decaying\/} exponential, 
\begin{equation}
\exp\biggl[ -{g\phi_0^ 2\over 2M} (t-t_0) \biggr] 
= \sum_{k=0}^\infty {(-1)^k\over 2^k k!} {g^k\over M^k} \phi_0^{2k} (t-t_0)^k .
\end{equation}
Knowing whether these corrections to the equation of motion can be organized into such a sum requires knowing the detailed coefficients of all of these terms generated by the loop graphs.

Or, to take a less complicated example, consider a damped harmonic oscillator whose motion is described by, $x(t)=x_0 e^{-\gamma t}e^{i\omega t}$.  If we had treated this system in the limit where $\gamma t$ is very small, we should have arrived at a perturbative solution, $x(t)=x_0 e^{i\omega t}[1-\gamma t + {1\over 2}\gamma^2t^2 + \cdots]$.  Had we then tried to extrapolate forward, without knowing that the powers of $\gamma t$ sum to $e^{-\gamma t}$, we should have falsely concluded that the solution diverges at late times.

The structures of these corrections are strongly reminiscent of similar growing terms that occur in the treatment of relaxation phenomena in quantum field theory.  In \cite{Boyanovsky:2003ui} it was shown that these terms could be organized into a decaying exponential function through a technique called the {\it dynamical renormalization group\/}.  We shall explore the resolution of these integral corrections in later work \cite{later}.  If they too sum to an decaying exponentially function, then these apparently dangerous terms would in fact contribute far more mildly to the one-point function than the transient effects which decay only as $(t-t_0)^{-3/2}$.  Thus, the net effect of the integral corrections could be vanishingly small by the times appropriate for the effective theory, and these corrections might require no new operators of the effective theory.

\subsection{More loops, $N$-point functions, and further generalizations}

While we have examined one class of graphs---the one-loop, $n$-vertex corrections to the one-point function---we have not exhaustively evaluated all possible loop corrections to all possible $N$-point functions.  While such a comprehensive treatment is beyond what we mean to cover here, we ought to mention at least a few observations about such higher order processes.

One might worry whether graphs with more loops produce the same $(t-t_0)^{-3/2}$ decay that we found to occur universally for the one-loop graphs.  We have evaluated one simple two-loop graph and found that it does produce this same nonanalytic $-{3\over 2}$ decay, although from a perturbative perspective it is suppressed by a further power of the coupling $\tilde \lambda$ relative to the one-loop contribution.  Further work is needed to understand completely what an arbitrarily complicated virtual process in the full theory, involving the heavy fields, implies for the low energy theory.

For the one-point function to provide nontrivial information about the effective theory, the light field needs to have a classical expectation value.  This step was done more for convenience than out of necessity.  We have also analyzed the two-point function, without a background field, $\phi(t)=0$, and found that more or less the same structures arise.  For the two-point function, the integrations by parts move the time-derivatives from the heavy loop to the external legs of the light field, which are now genuine propagators rather than simply powers of the background field.  Compared with even just the two-point function, the one-point function is far simpler to analyze.  For a general $N$-point function, we are not free to amputate all of the external legs in the Schwinger-Keldysh framework; moreover, for the two-point function and beyond, external momenta can flow through the loop, which means that we must perform both a time-derivative and a small external momentum spatial expansion of the loop kernel simultaneously.

\section{Conclusions and further directions}
\label{sec:conc}

In this article we have explored the construction of effective field theories for time-dependent, nonequilibrium systems. ÊOur basic tool for analyzing the possible structures that could appear in an effective theory was to take a top-down approach, starting with a simple theory with two scalar fields, one heavy and one light, and to demand that the effective theory composed only of the light field should reproduce the observables in a late time and low energy limit. Ê

In particular, we have calculated the equation of motion for the expectation value of the light field which includes the leading virtual effects of the heavy field. ÊWe started the evolution at a finite initial time and chose the initial state to be the vacuum of the free theory. ÊThe resulting equation of motion includes the usual local terms that would have been expected from a local, Poincar\'e-invariant action for the effective field theory, but we found that further terms are present which cannot arise from a local effective field theory.  Among them are some with an explicit time dependence, while others contain time integrals of powers of the light field. ÊIn the limit where the initial time is taken to the infinite past, and the coupling is turned on adiabatically, the nonlocal terms all vanish and a local effective theory is recovered.

If it appears that the action that we found is not what one would have expected of a theory when the heavy and light sectors are decoupled, which would have produced only the set of local operators, it is because our example essentially contains excitations of the heavier field.  The initial state, which was chosen to be the vacuum for the {\it free\/} Lagrangian of both fields, is not an eigenstate of the interacting theory.  So from the very start, we are not in an equilibrium state.  Once we have opened up the system thus, the evolution no longer needs to be unitary.  In our example, the nonlocal terms correspond to the annihilations of the excitations of the heavy field contained in our initial state.

Another way of seeing the origin of this nonlocality is to extend the evolution back to $-\infty$, and to reproduce the original setting by introducing the interaction as ${1\over 2}g\Theta(t-t_0)\chi^2\Phi^2$, which explicitly breaks the time-translation invariance.  We can do so since our initial states are the free vacuum states (we are here neglecting the self-interactions of the field), and by having the free theory prior to $t_0$ the fields remain in their vacua.  Of course this trick is a heuristic one, since all that we need to know is the action and the initial state:  everything prior to $t_0$ is irrelevant for the subsequent evolution.  But seen as a term that turns on suddenly---and being a step-function it is sudden from the perspective of the heavy field too---we are introducing a finite energy density into the system at $t_0$ which is spread over all momentum modes, including heavy ones.

So the signature of some residual nonlocality as the heavy fields annihilate and the system relaxes is not in itself surprising.  But what we have found is that among some of these nonlocal effects there is a universal time-dependence, a characteristic $-{3\over 2}$ decay inherited from the structure of the heavy loop.  The existence of these decaying operators means that the rules change slightly.  For a local operator suppressed by $M^{-n}$, if we perform an experiment with a maximal energy of $E$ and a precision of $\varepsilon$, then our predictions should still take account of all the operators where $(E/M)^n \gtrsim \varepsilon$.  But for a class of the nonlocal operators, we must also keep those where 
$$
{1\over [M(t-t_0)]^{(3/2)+n'}} \biggl( {E\over M} \biggr)^n \gtrsim \varepsilon 
\qquad\quad 
n,n'=0,1,2,3, \cdots .
$$
So the set of nonlocal operators that we need varies with time.  If we wait long enough---and what we mean by `long enough' depends on the energies and precision of our experiment---then we do not need any of them.  If we make optimal use of our experiment, so that $(t-t_0)^{-1}\sim E$, then all we have done is to separate out a specific power of $E$ from the usual rule.  However, it is a nonanalytic power, and that would not have been expected of an entirely local effective theory.  The time-dependent decay is slow enough that even when we are below the threshold for exciting the heavier fields directly from the light fields, there is a long enough tail in the annihilation that they are still producing light fields during the regime $1\ll M(t-t_0)\ll \infty$, which might be observable depending on the sensitivity of our experiment.  The relaxation is not `instantaneous' according to the resolution of our detector.

This example is fairly specific, but we expect that there are general lessons which we can learn from it.  Had we started in a more general nonequilibrium state, as long as some heavy modes are excited, we should expect to find similar structures:  nonlocal operators, decaying coefficients.  We must, of course, not include {\it too large\/} an energy density in the initial state if we wish to be able to apply an effective treatment.  From the alternative perspective where we turn on the coupling, we could equivalently consider more general time-dependences, ${1\over 2} g(t)\chi^2\Phi^2$; but cases where $g(t)$ has continually occurring features that are appreciable on intervals $\Delta t\sim M^{-1}$ would lie outside an effective treatment altogether.

In the opposite regime, we could consider a time-dependent system in which the local effective theory remains always applicable.  For instance, considering a time-dependent coupling again of the form ${1\over 2} g(t)\chi^2\Phi^2$, if we turn on $g(t)$ sufficiently slowly---though not necessarily adiabatically---so that we are only exciting modes below $M$, then we should not expect to see the sorts of nonlocal effects that we found when we excite the heavy modes directly.  Of course, the system is still a time-evolving, nonequilibrium one, where energy can be exchanged amongst different momentum modes of the light field.  We saw an example of this sort of time-dependence in the equation of motion for the light field back when we introduced the tadpole method.  But as long as none of the heavy fields are excited directly, their existence should only enter the theory through local operators of the light field.

These conjectures about the form of the effective theory in the presence of more general time-dependent structures still need to be checked rigorously.  And even for the comparatively simple example that we have analyzed here, a few open questions remain which also must be addressed, such as the fate or role of the terms which appear at order $g^3$ and above that contain integrals of powers of the light field \cite{later}.  One route for analyzing the structure of the effective theory more directly, at the level of the path integral rather than through the tadpole method used here, would be to express the virtual effects of the heavy field through an {\it influence functional\/} \cite{Feynman:1963fq}.  By appropriately expanding this influence functional in the limit where the momenta entering it are small compared to $M$, we could see what structures appear in the path integral.  From this we could then try to infer the appropriate generating function for the effective theory.  The appearance of time-dependent structures in the high energy action, such as the coupling here that `turns on' suddenly or some more general coupling with an ongoing time-dependence, represent vertices that do not conserve energy.  Analyzing these vertices in the time-frequency domain, we can see exactly which frequencies of which fields are being excited by a particular time-dependent operator.

Ultimately we should like to apply what we have learned to systems where there is a natural time-dependence, especially in instances where the higher energy theory might be unknown.  In inflation, there is a rather dramatic time-dependence produced by the expansion of the space-time.  The existence of the oscillations in the nonlocal, decaying terms might then be able to produce interesting features in the spectrum of primordial fluctuations, for example.  Some effort has already been made to apply the ideas of effective field theories to inflationary backgrounds \cite{Weinberg:2005vy,Weinberg:2008hq,Cheung:2007st,Senatore:2009cf}.  And some works too have specifically examined the influence of a heavy field on the correlation functions of the inflaton \cite{Achucarro:2012sm,Achucarro:2010da,Shiu:2011qw,Avgoustidis:2012yc}, using classical field theory.  The picture that is emerging is a very intriguing one.  It is clear that there is still much to be learned from effective theories in time-dependent settings.

\acknowledgments
R.H.~would like to thank the U.S. Department of Energy for partial support through grant DE-FG03-91-ER40682, and A.R.~would like to thank the National Aeronautics and Space Administration for support through NASA grant 22645.1.1110173. H.C.~is grateful for the support of the physics department of Carnegie Mellon University.  We should also like to thank Ira Rothstein, whose questions were the inspiration for this work, Cliff Burgess, John Donoghue, and Lorenzo Sorbo for useful discussions.

\end{document}